\documentclass[]{IEEEtran}
\usepackage{mathrsfs}
\usepackage{amsfonts}
\usepackage{graphicx,cite,epsfig,amssymb,amsmath}
\usepackage{color,xcolor}
\definecolor{red}{rgb}{1.00, 0.00, 0.00}
\usepackage{pifont}
\usepackage{stmaryrd}
\usepackage{setspace}
\usepackage{subfigure}
\usepackage{cite}
\usepackage{array}
\usepackage{float}
\usepackage{verbatim}
\usepackage{multirow}
\usepackage{booktabs}
\usepackage{makecell}
\usepackage[linesnumbered, ruled]{algorithm2e}
\usepackage{epstopdf}
\usepackage{mathtools}
\usepackage{diagbox}
\usepackage{cases}
\usepackage{amsthm}
\usepackage[colorlinks,
linkcolor = black,
anchorcolor = black,
citecolor = black]{hyperref}
\usepackage{textcomp}

\providecommand{\algorithmname}{Algorithm}
\floatname{algorithm}{\protect\algorithmname}
\newcommand{\bm}[1]{\mbox{\boldmath{$#1$}}}
\newtheorem{thm}{Theorem}

\newtheorem{prop}{Proposition}
\usepackage{arydshln}
\allowdisplaybreaks[4]

\newtheoremstyle{noparens}%
  {}{}%
  {\itshape}{}%
  {\bfseries}{.}%
  { }%
  {\thmname{#1}\thmnumber{ #2}\mdseries\thmnote{ #3}}
\theoremstyle{noparens}

\ifodd 1

\else

\fi

\begin{document}
	
\title{Sensing Framework Design and Performance Optimization with Action Detection for ISCC}
\author{\IEEEauthorblockN{Weiwei Chen, Yinghui He,  Guanding Yu, \IEEEmembership{Senior Member,~IEEE}}, Jianfeng Wang, and Haiyan Luo\\

\thanks{
Manuscript received 13 October 2024; revised 05 March 2025; accepted 27 April 2025. (\emph{Corresponding author: Yinghui He})

W. Chen and Y. He are with the College of Information Science and Electronic Engineering, Zhejiang University, Hangzhou 310027, China. e-mail: \{22331145, 2014hyh\}@zju.edu.cn.

G. Yu is with the State Key Laboratory of Ocean Sensing, Zhoushan 316021, China, and also with the College of Information Science and Electronic Engineering, Zhejiang University, Hangzhou 310027, China. e-mail: yuguanding@zju.edu.cn.

J. Wang is with Lenovo Research, Lenovo Group Ltd., Beijing 100094, China. e-mail: wangjf20@lenovo.com.

H. Luo is with Lenovo (Shanghai) Information Technology Co., Ltd, Shanghai 201203, China. e-mail: luohy7@lenovo.com.}}


\maketitle

\begin{abstract}
Integrated sensing, communication, and computation (ISCC) has been regarded as a prospective technology for the next-generation wireless network, supporting human-centric intelligent applications.
However, the delay sensitivity of these computation-intensive applications, especially in a multi-device ISCC system with limited resources, highlights
the urgent need for efficient sensing task execution frameworks. 
To address this,
we propose a resource-efficient sensing framework in this paper. 
Different from existing solutions, it features a novel action detection module deployed at each device to detect the onset of an action. 
Only time windows filled with signals of interest are offloaded to the edge server and processed by the edge recognition module, thus reducing overhead. 
Furthermore,
we quantitatively analyze the sensing performance of the proposed sensing framework and formulate a sensing accuracy maximization problem under power, delay, and resource limitations for the multi-device ISCC system.
By decomposing it into two subproblems,
we develop an alternating direction method of multipliers (ADMM)-based distributed algorithm. It alternatively solves a sensing accuracy maximization subproblem at each device and employs a closed-form computation resource allocation strategy
at the edge server till convergence. Finally, a real-world test is conducted using commodity wireless devices to validate the sensing performance analysis. 
Extensive test results
demonstrate that our proposal achieves higher sensing accuracy under the limited resource compared to two baselines.

\end{abstract}


\section{Introduction}
The next-generation
wireless network is envisioned to push forward an era of Intelligence of Everything (IoE), providing various intelligent services with smart ambient sensing~\cite{cui2021integrating}.
To achieve this vision, integrated sensing and communication (ISAC)~\cite{isac1,isac2,isac3} has been recognized as one of key technologies by IMT-2030~\cite{ITU2023}. 
By exploring the potential of using the same communication hardware and spectrum for sensing, ISAC endows the wireless network with native high-performance sensing capabilities while incurring minimal overhead~\cite{zhang2022practical}.
This enables future wireless networks to not only support traditional sensing applications (e.g., localization and tracking), but also extend to human-centric sensing applications (e.g., action recognition and respiratory monitoring)~\cite{li2022human,10556745}, thereby facilitating advanced services, such as smart homes, extended reality, and remote healthcare~\cite{saad2019vision}.
To address computational demands of the human-centric sensing applications, integrated sensing, communication, and computation (ISCC) has been proposed by jointly considering ISAC and mobile edge computing (MEC)~\cite{zhao2022radio,li2023over,zhao2024multi,liu2024joint,wan2024ris}.

To realize human-centric sensing, existing approaches typically employ Internet of Things (IoT) sensing transceivers
to consistently transmit and receive sensing signals, 
with the collected data streams segmented into series over time windows and further fed into computation-intensive machine learning (ML) algorithms~\cite{ullmann2023survey}. 
However, such a straightforward framework would result in significant resource waste, as continuous data transmission and ML-based processing are often unnecessary, particularly when no active sensing targets are present. This issue is exacerbated in multi-device ISCC systems, where multiple devices compete for limited communication and computation resources for their own sensing tasks. 
To address it, we aim to propose a resource-efficient sensing framework specifically designed for multi-device ISCC systems. 

\subsection{Related Work}

With the rapid advancement of artificial intelligence (AI), human-centric sensing tasks have become a research hotspot. To support these tasks, ISCC has emerged as a promising technology, which jointly considers three coupled processes essential for task execution: sensing (i.e., data acquisition), communication (i.e., data transmission), and computation (i.e., data processing). ISAC serves as one fundamental technology for ISCC, and it aims to realize high-performance sensing using communication signals without additional hardware and spectrum resources.
Existing research on ISAC can generally be classified into four categories: resource allocation~\cite{isac-res1,isac-res2}, interference management~\cite{he2022ris,inter-mana3}, signal processing~\cite{isacsignal2,yu2023active,he2023sencom}, and performance analysis~\cite{zhuo2023performance}.
For instance, authors in~\cite{he2023sencom} investigated the calibration of channel state information (CSI) from general WiFi communication packets, providing high-quality and adequate CSI
for upper-layer sensing applications.
Nevertheless, the computation process of sensing data should also be considered, as sensing tasks, such as activity recognition, generally rely on well-trained deep neural networks (DNNs), which demand substantial computational resources for real-time inference~\cite{hua2023edge}. To address this issue, MEC, another fundamental technology for ISCC, provides a feasible solution by offloading sensing data to edge servers for further processing~\cite{mao2017survey,zhang2018joint,xiao2022multi}.  The studies of MEC for AI-driven tasks have primarily focused on reducing overhead from the perspective of joint communication and computation. For example,~\cite{multi1} proposed a joint communication and computation resource allocation algorithm for a multi-user edge inference system, leveraging batching and early exiting to reduce latency and improve throughput. 
This idea was further extended to multi-modal learning systems, as explored in~\cite{multi2}.

Nevertheless, the techniques mentioned above focus only on partial aspects of the task execution processes, thereby constraining their performance potential, particularly in resource-constrained scenarios~\cite{zhu2023pushing}. Consequently, the core challenge in designing ISCC systems for AI-driven tasks lies in how to allocate resources by jointly considering sensing, communication, and computation processes. Several existing studies have proposed their solutions from different perspectives~\cite{wen2023task,wen2023taskair,wang2023device,du2024integrated}.
Specifically,~\cite{wen2023task} mathematically characterized the impact of sensing, communication, and computation on sensing accuracy using a novel metric, namely discrimination gain. Then, a task-oriented joint resource allocation scheme was proposed to enhance the sensing performance. Subsequently, a multi-device ISCC system was proposed in~\cite{wen2023taskair}, with over-the-air computation (AirComp) used to perform the ensemble inference to reduce the latency. Additionally, the transmit precoder and receive beamforming were jointly optimized to maximize the inference accuracy.
To address privacy concerns during the inference process,~\cite{wang2023device} proposed a joint local and ensemble inference scheme for AirComp-assisted ISCC systems. 
Moreover, with the achievements of 
federated learning (FL) techniques,~\cite{du2024integrated} developed an ISCC-FL framework for target sensing, 
and proposed a joint beamforming and device scheduling strategy to enhance the performance of FL within ISCC systems.

\subsection{Main Contributions}
The above ISCC works related to human-centric sensing applications 
primarily focus on resource allocation
to improve the accuracy~\cite{wen2023task,wen2023taskair}, or avoiding the privacy leakage~\cite{wang2023device,du2024integrated}.
However, directly implementing ML-based algorithms to all sensing time series would lead to high communication and computation overhead, since sensing areas may lack sensing targets, or sensing targets may remain static for a long time. 
To this end,~\cite{he2023integrated} proposed an action detection module for detecting whether there is an action within the current time window, while overlooking the continuity
of sensing tasks over time. This would lead to time windows 
containing only partial sensing signals of interest 
being fed to the ML algorithms, thus wasting resources and impeding the recognition accuracy. To address this, we propose a resource-efficient
sensing
framework. Different from~\cite{he2023integrated}, 
it contains a novel action detection module at the IoT device that aims to find the onset of an action, and only the time window fully filled with the signal of interest would be uploaded to the edge server for further processing.
Moreover, considering the limitations of the computation resource in a practical multi-device ISCC system, we further develop an alternating direction
method of multipliers (ADMM)-based distributed algorithm to maximize the sensing performance.
The main contributions of this paper can be summarized as follows.
\begin{itemize}
    \item We propose a resource-efficient sensing framework consisting of a novel action detection module deployed at each device and an edge recognition module deployed at the BS. The action detection module detects the onset of an action by comparing the power difference of high-frequency components between two consecutive time windows with a threshold, avoiding unnecessary subsequent transmission and computation. 
    \item We quantitatively analyze the sensing performance for the proposed sensing framework, in terms of sensing accuracy, power consumption, and sensing delay. 
    A sensing accuracy maximization problem is further formulated under the constraints of power, delay, and computation resource.    
    \item To solve this problem, we propose an ADMM-based distributed algorithm with minimal information exchange. It contains a sensing accuracy maximization algorithm at each device by optimizing the parameters in the sensing framework and a closed-form computation resource allocation strategy at the edge server. 
    \item To evaluate the performance of the proposed sensing framework and algorithm, we conduct a real-world test
    using commodity wireless devices. 
    Extensive test results validate the correctness of the sensing model analysis for the proposed framework and 
    verify the effectiveness of the proposal compared to two baselines. 
\end{itemize}

The rest of this paper is organized as follows. 
Section~\ref{sec:sys_model} introduces the ISCC system and outlines the workflow of the proposed sensing framework. Section~\ref{sec:perfrom} analyzes the performance of the sensing framework and formulates a sensing accuracy maximization problem. An ADMM-based algorithm is proposed
in Section~\ref{sec:solve}. 
Section~\ref{sec:test} presents test results for model validation and performance verification, and the whole paper is concluded in Section~\ref{sec:Conclusion}.

\section{System Model and Sensing Framework}\label{sec:sys_model}
In this section, we introduce the considered ISCC system, and the workflow of the proposed sensing framework.

\subsection{ISCC System and Sensing Framework} 
\begin{figure}[t]
	\setlength{\abovecaptionskip}{8pt} 
		\centering		\includegraphics[width=0.98\linewidth]{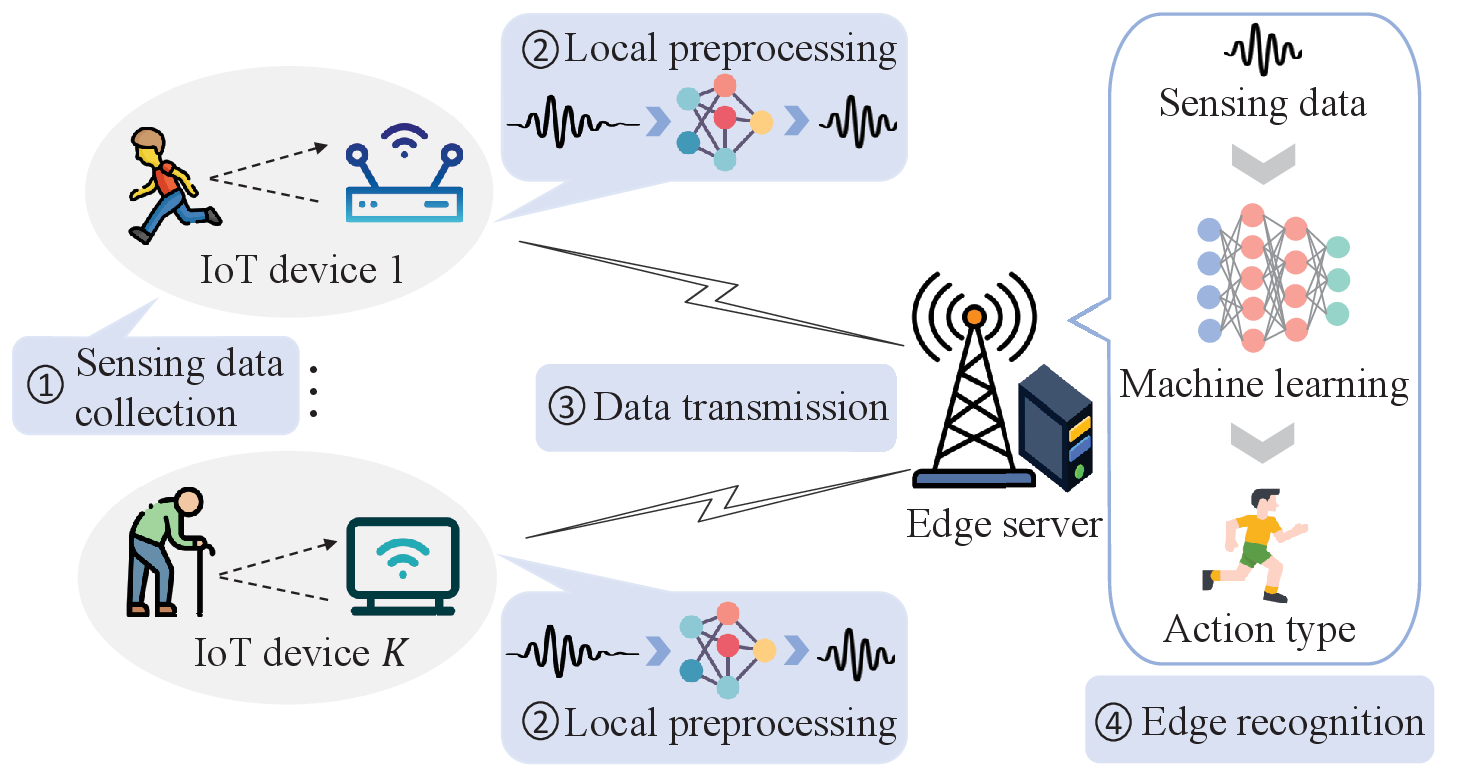}
		\vspace{-2ex}
		\caption{The ISCC system and the proposed sensing framework.}
		\label{fig:sys}
  \vspace{0.4em}
       \centering
	\includegraphics[width=0.92\linewidth]{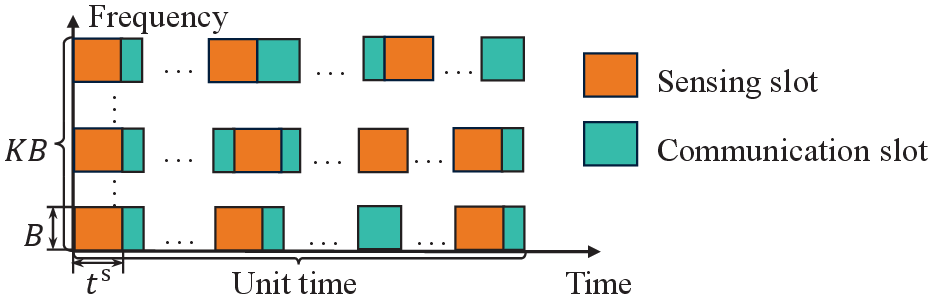}
		\vspace{-2ex}
		\caption{Time and frequency resource allocation in the ISCC system.}
		\label{fig:frame}
        \vspace{-3ex}
\end{figure}

As illustrated in Fig.~\ref{fig:sys}, we consider a multi-device ISCC system comprising a BS equipped with an edge server and $K$ IoT devices. Each IoT device has its own sensing task: 
detect the sensing target in a sensing area and further recognize the action of the target detected. 
To this end, all IoT devices are equipped with dual-functional 
transceivers, allowing them to serve both sensing and communication purposes\footnote{We assume that each IoT device is equipped with one transmit antenna and one receive antenna, which aligns with the constraints of many practical IoT applications. Additionally, our proposed framework can be extended to the case with multi-antenna devices.}. 
Specifically, the entire sensing framework from data collection to outputting action types can be divided into four steps:
\begin{itemize}
    \item[1)] Sensing data collection: Each IoT device transmits wireless sensing signals continuously toward its designated sensing area 
    and collects the corresponding echo signals.     
    \item[2)] Local preprocessing: Each device preprocesses the collected sensing signals. This step also contains a proposed novel action detection module for avoiding waste of resources, which will be detailed in Section~\ref{sec:action_detection}.
    \item[3)]Data transmission: Due to the limited computation resource at the IoT device, the preprocessed sensing data is further uploaded to the edge server via wireless links for action recognition.
   \item[4)] Edge recognition: The edge server applies ML-based algorithms to recognize the action type of each sensing target with the received sensing data.
\end{itemize}

To avoid inter-user interference, particularly among sensing signals of different devices, that could corrupt the detection of the target object, the BS adopts the orthogonal frequency-division multiple access (OFDMA) method for multi-device access.
Each IoT device occupies one resource block for sensing and communication with the bandwidth of each resource block being $B$. 
Moreover, a time-division manner is employed to flexibly switch between sensing and communication modes, as shown in Fig.~\ref{fig:frame}\footnote{This time-division approach is compatible with existing IoT systems. For example, an IoT device can incorporate an OFDM radar for sensing and a 5G communication module for data transmission. Additionally, the framework can be extended to a
millimeter-wave (mmWave) system, where an mmWave radar is employed for sensing and an mmWave communication module facilitates data transmission.}. In the sensing mode, IoT device $k$ transmits sensing signals at a sampling rate $F_k^{\mathrm{s}}$ and receives echoes to obtain sensing data.
Let $t^{\mathrm{s}}$ denote the duration of 
transmitting one sensing signal and receiving the corresponding echo
per unit time, and thus the total proportion of time slots occupied by sensing is $t^{\mathrm{s}} F_k^{\mathrm{s}}$. Note that $t^{\mathrm{s}} F_k^{\mathrm{s}}$ should be less than 1, and the remaining proportion $(1-t^{\mathrm{s}} F_k^{\mathrm{s}})$ is used for communication.

\subsection{Sensing Basic and Motivation} \label{sec:sen_pre}

During the data collection phase, each IoT device transmits an orthogonal frequency-division multiplexing (OFDM) radar~\cite{barneto2019ofdm} signal within the allocated resource block to obtain the sensing data. 
Specifically, each resource block contains $N$ subcarriers, and the transmit signal at subcarrier $n$ is denoted by $x_{k,n}[l]$ for the $k$-th IoT device at time index $l$. After being reflected by the sensing target,
the signal received at the IoT device is $y_{k,n}[l]$. Then, the CSI can be calculated by $h_{k,n}[l]=\dfrac{y_{k,n}[l]}{x_{k,n}[l]}$. Since the CSI contains the information of the sensing target, it can be used for sensing. 
To realize the sensing for an action, each IoT device needs to collect CSI over a time window, denoted by $T^\mathrm{s}$, and then constructs a CSI matrix as $\bm{H}_k[l] =[\bm{h}_{k,1}[l], \bm{h}_{k,2}[l],  \cdots, \bm{h}_{k,N}[l] ]^H \in \mathbb{C}^{N\times( T^{\mathrm{s}}F^{\mathrm{s}}_k)}$ at time $l$, where $\bm{h}_{k,n}[l]$ is the CSI vector consisted of last $T^{\mathrm{s}}F^{\mathrm{s}}_k$ collected CSI samples for the $n$-th subcarrier, i.e., $\bm{h}_{k,n}[l] = \left[h_{k,n}[l-T^{\mathrm{s}}F^{\mathrm{s}}_k+1], \cdots, h_{k,n}[l] \right]^H \in \mathbb{C}^{( T^{\mathrm{s}}F^{\mathrm{s}}_k)\times 1}$, and $(\cdot)^H$ denotes the operation of conjugate transpose. 

\begin{figure}[t]
	\setlength{\abovecaptionskip}{8pt} 
	\centering
    \includegraphics[width=0.9\linewidth]{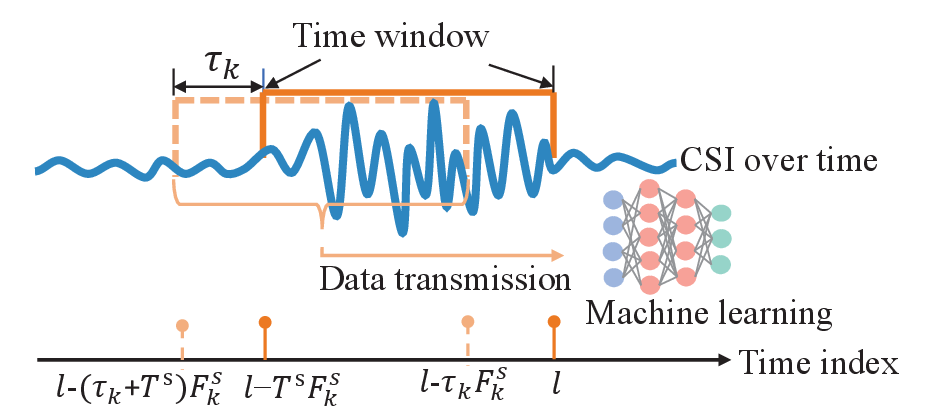}
	 \vspace{-2ex}
	\caption{Illustration of time window and time step.} 
	\label{fig:sen_win}
 \vspace{-3ex}
\end{figure}

Processing a CSI matrix every time a new CSI sample is received would incur significant overhead. To reduce the overhead, we can perform the subsequent processing after a time step, denoted by $\tau_k$ (in second) for the $k$-th IoT device, as shown in Fig.~\ref{fig:sen_win}. In traditional sensing systems, IoT devices continuously send collected sensing data to the edge server every $\tau_k$ time interval, corresponding to $\tau_k F_k^{\mathrm{s}}$ samples over time. In other words, $l$ only could be $0, F_k^{\mathrm{s}}\tau_k,2F_k^{\mathrm{s}}\tau_k, \cdots $.
However, in some scenarios, such as in smart home scenarios, sensing areas may lack sensing targets, or sensing targets may remain stationary for a quite long time. 
Consequently, continuously transmitting data to the edge and running sensing algorithms would incur a significant waste of communication and computation resources. Meanwhile, it is widely acknowledged that the movement of objects within the area of interest would cause CSI to fluctuate over time remarkably, resulting in an increase in the power of high-frequency components of the CSI over time~\cite{wifi-csi}.
Thus, authors in~\cite{he2023integrated} filtered out static targets by comparing the power of the CSI over time $T^{\mathrm{s}}$ to a preset threshold, thereby conserving edge computation resource.
However, the proposed method in~\cite{he2023integrated} does not consider the impact of the time step and assumes that the CSI collected in each time window are all from signals of interest caused by the movement of the sensing target, overlooking the continuity of the sensing signals.

 \begin{figure}[t]
	\vspace{-4ex}
	\centering
	\includegraphics[width=0.75\linewidth]{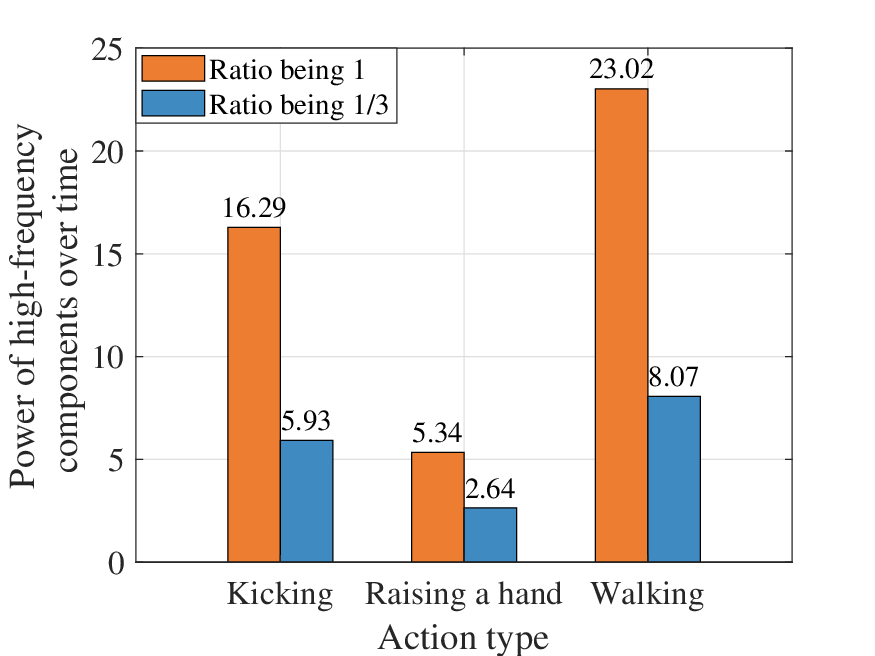}
	\vspace{-2ex}
	\caption{The power of high-frequency components over time for three different actions under varying ratios.}
	\label{fig:ener_type}
    \vspace{-1em}
\end{figure}

\begin{figure*}[t]
\setlength{\abovecaptionskip}{8pt} 	
	\centering
    \subfigure[The first phase.]{
		\centering
		\includegraphics[height=0.12\linewidth,trim=0 3 0 0,clip]{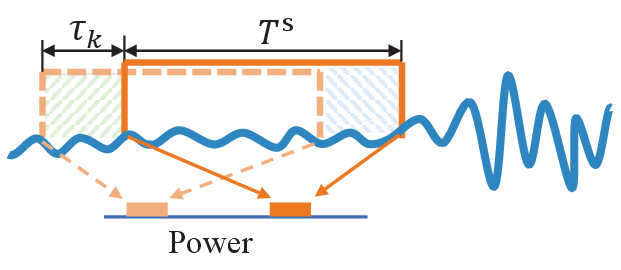}
	}
    \subfigure[The second phase.]{
		\centering
		\includegraphics[height=0.12\linewidth,trim=0 3 0 0,clip]{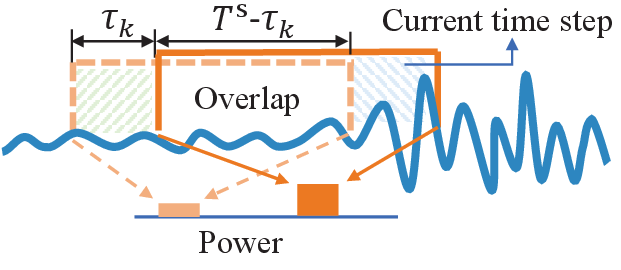}
         \label{fig:sec-phase}
	}
     \subfigure[The third phase.]{
		\centering
		\includegraphics[height=0.12\linewidth,trim=0 3 0 0,clip]{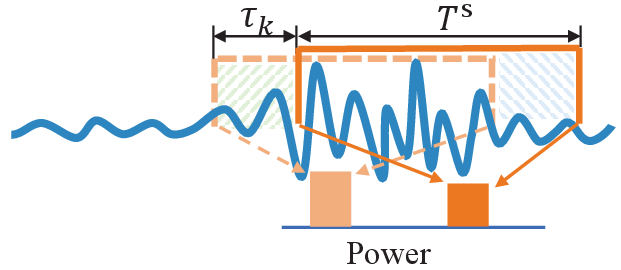}
	}
	  \vspace{-2ex}
	\caption{Power difference of high-frequency components in three phases.} 
   \vspace{-3ex}
	\label{fig:en_diff scenario}
\end{figure*}

As we can imagine, only part of the CSI signals contain the information associated with the movement within a time window. The ratio, which is defined as the proportion of the signal of interest within the time window, varies randomly due to the randomness of the movement’s start time, as illustrated in Fig.~\ref{fig:sen_win}.
This phenomenon would impede the detection in~\cite{he2023integrated} that only leverages the power within one time window, since the power of high-frequency components would be different when the ratio is different, even if the action is the same. 
To verify this, we collect CSI over time for different actions, and the power of high-frequency components is illustrated in Fig.~\ref{fig:ener_type}. 
Specifically, we showcase the power of high-frequency components for three different actions (i.e., kicking, raising a hand, and walking) under varying ratios. Following the method in~\cite{he2023integrated}, the threshold should be set below 5.34 to avoid misdetection. However, CSI signals of kicking and walking with the ratio being 1/3 would pass the detection under this threshold, and performing action recognition on them would cause a waste of resources and degrade the recognition accuracy since the sensing information contained in them is not enough for distinguishing generally.
To address this issue, we aim to design a novel action detection module deployed at each IoT device. 
Different from~\cite{he2023integrated}, the proposed module aims to find the onset of an action.

\subsection{Action Detection Module}\label{sec:action_detection}

As described in Fig.~\ref{fig:sen_win}, the ratio of the signals of interest gradually increases as the time window progresses, leading to a gradual rise in the power of high-frequency components. 
Leveraging this rationale, we propose a novel action detection module by 
judging the power variation
to determine whether the time window is filled with signals of interest.
Specifically, we first select the CSI vector of the first subcarrier from $N$ subcarriers\footnote{Here, we take only the first subcarrier mainly for a low computational complexity.
To further improve the detection accuracy, all $N$ subcarriers can be utilized for the action detection module with the help of a voting classifier.}. Then, we convert the CSI vector from the time domain to the frequency domain in order to obtain the power of high-frequency components. 
To this end, at time index $l$, we perform a discrete Fourier transform (DFT) on the CSI vector, as 
\begin{equation}
	\!\!\!W_k[f,l] \!=\! \frac{1}{\sqrt{T^{\mathrm{s}}F^{\mathrm{s}}_k}}\!\sum_{m=l\!-\!T^{\mathrm{s}}F^{\mathrm{s}}_k\!+\!1}^{l}\!\!\!h_{k,1}[m] \exp\left(\frac{-j 2\pi f m}{T^{\mathrm{s}}F^{\mathrm{s}}_k}\right).
\end{equation}              
Let $[F^{\ell}, F^{\mathrm{u}}]$ denote the range of high frequency. The power of high-frequency components within the time window at time $l$ can be calculated as the summation of squares of the DFT $W_k[f,l]$ within the range $\left[\lfloor F^{\ell}T^{\mathrm{s}}\rfloor, \lceil F^{\mathrm{u}}T^{\mathrm{s}} \rceil\right]$, as
 \begin{equation}
		E_k[l] = \sum_{f= \lfloor F^{\ell}T^{\mathrm{s}}\rfloor }^{ \lceil F^{\mathrm{u}}T^{\mathrm{s}}\rceil } \left| W_k[f,l] \right|^2.
\end{equation}
Thus, the power can be calculated as
\begin{equation}
    P_k[l] = \frac{1}{T^{\mathrm{s}} F^{\mathrm{s}}_k} E_k[l].
\end{equation}

We assume that there is no sensing target in the sensing area at the beginning of sensing. As time progresses, there are three phases during the sensing, as shown in Fig.~\ref{fig:en_diff scenario}. In the first phase, there is no sensing target, and the power difference of high-frequency components between two consecutive time windows is minor since it is caused by the noise. In the second phase, part of the time window is filled with the signal of interest, and the power increases as time progresses with an obvious power difference between two consecutive time windows. In the third phase, the power may stabilize or begin to drop, which is determined by the duration of the sensing target's action. 
Thus, we can compare the power difference with a preset threshold, denoted by $\eta_k$, to detect and timestamp the onset of the second phase, which is also the onset of the action.
Specifically, the power difference at time index $l$ for the $k$-th IoT device can be calculated as 
\begin{equation} \label{eq:delta_p}
    \Delta P_k[l]= P_k[l]-P_k[l-\tau_k F^{\mathrm{s}}_k].
\end{equation}
By comparing the power difference $\Delta P_k[l]$ to the threshold $\eta_k$, we can determine whether the movement of the sensing target has started. If $\Delta E[l]$ is higher than $\eta_k$, the movement is assumed to have occurred
between the time index interval $[l\!-\!\tau_kF^{\mathrm{s}}_k, l]$. 
According to~\cite{possion}, we assume that the occurrence of the movement follows a Poisson process, and then the average onset is $l\!-\!\tau_kF^{\mathrm{s}}_k/2$.
Thus, the subsequent collected CSI matrix within the time window of $[l\!-\!\tau_kF^{\mathrm{s}}_k/2, l\!-\!\tau_kF^{\mathrm{s}}_k/2\!+\!T^{\mathrm{s}} F^{\mathrm{s}}_k]$ is further input into the next step, i.e., uploaded to the edge server.
Otherwise, the IoT device waits for the next time window and continues performing detection. 

It is evident that both the time step and the threshold are critical for the action detection module. If the threshold is set too low, the onset of movement may be detected incorrectly. 
In this case, the sensing signals in some time windows belonging to the first phase would be mistakenly transmitted to the edge, wasting both communication and computation resources. Conversely, if the threshold is set too high, the onset of movement may be missed, leading to a decrease in the recognition accuracy. Regarding the time step, it directly influences the ratio of the signal of interest within the time window at the start of the second phase, as shown in Fig.~\ref{fig:sec-phase}. A larger time step enhances the power difference, making it easier to detect the onset of an action. However, the precision of the onset, i.e., the difference between the actual onset and the estimated onset, would increase with the time step since we only detect whether the onset is within the time step but cannot precisely find the onset. This would further influence the accuracy 
of the edge recognition.
Additionally, the sampling rate also affects the accuracy. A higher sampling rate brings more information about the sensing target, thereby enhancing the accuracy. 
In Section~\ref{sec:perfrom}, 
we will explore the mathematical relationship among the threshold, time step, sampling rate, and sensing accuracy for further optimizing the performance of the proposed framework. 


\subsection{Data Transmission and Edge Recognition}
After undergoing the action detection, each IoT device uploads the CSI matrix that passes the detection 
to the edge server through the OFDMA method. For device $k$, the instantaneous data rate can be expressed as 
\begin{equation}
R_{k} = \frac{B}{N}\sum_{n \in \mathcal{N}_k} \log_2\left(1+ \frac{|H_{k,n}|^2 P_{k}^{\mathrm{t}}}{N \sigma^2}\right),
\end{equation}
where $\mathcal{N}_k$ and $P_{k}^{\mathrm{t}}$ represent the allocated subcarrier set and the transmission power for device $k$, respectively. Moreover, $H_{k,n}$ denotes the channel gain at subcarrier $n$ between device $k$ and the BS, and $\sigma^2$ is the noise power of each subcarrier. Since the time slot proportion allocated to communication is $1-t^{\mathrm{s}} F_k^{\mathrm{s}}$, the average data rate is $(1-t^{\mathrm{s}} F_k^{\mathrm{s}})R_k$.

After receiving the CSI matrix from each device, the edge server employs convolutional neural network (CNN)-based sensing
algorithms to realize action recognition and then feeds the result back to each device.

\section{Performance Analysis and Problem Formulation}\label{sec:perfrom}
In this section, we first analyze the sensing accuracy, power consumption, and sensing delay of our proposed sensing framework, and then formulate a multi-device sensing performance maximization problem.

\subsection{Sensing Accuracy} \label{sec:sensing}
We assume that each sensing task is an $I$-class action recognition task, where the first class (i.e., $i=1$) represents the static state and the remaining classes correspond to different action types. The sensing accuracy is related to two modules: the action detection module and the edge recognition module.

\subsubsection{Action Detection Module}
\begin{figure}[t]
	\setlength{\abovecaptionskip}{8pt} 	
	\centering
    \includegraphics[width=0.95\linewidth]{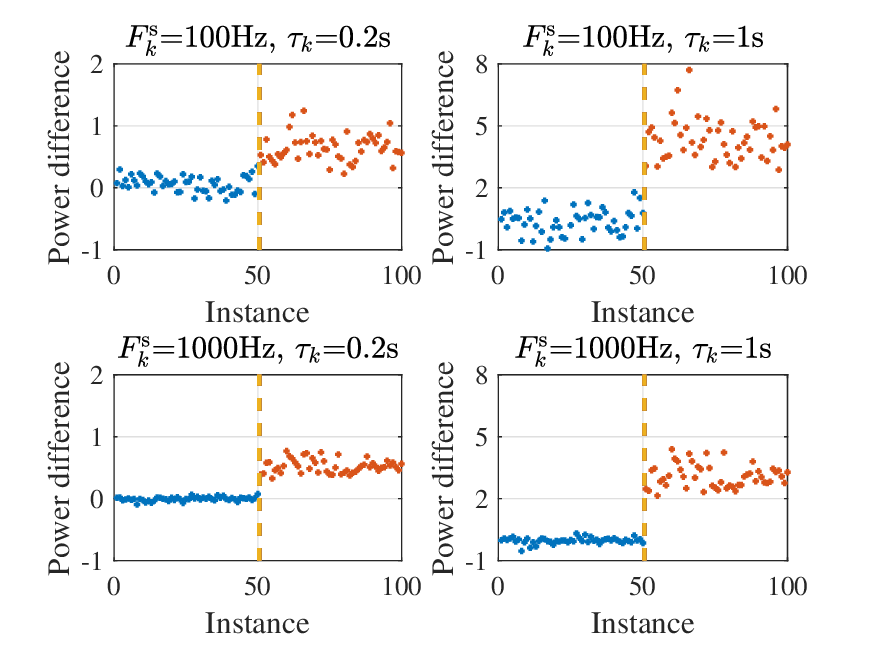}
	\vspace{-3ex}
	\caption{The power difference of high-frequency components for 100 instances under different sampling rates and time steps. The first 50 instances correspond to the first phase in Fig.~\ref{fig:en_diff scenario},
and the last 50 instances correspond to the second phase.} 
 \vspace{-4ex}
	\label{fig:energy_instance}
\end{figure}

The performance of this module can be described by false positive rate and miss rate~\cite{bai2020carin}.
Specifically, the former, denoted by $p^{\mathrm{l}}$, indicates the probability of incorrectly detecting the onset of an action when the time window is in static state,
while the latter, denoted by $p^{\mathrm{o}}_i,i \in \{2,3,\cdots,I \}$, refers to the probability of failing to detect the actual onset of the $i$-th type of action.  
As previously discussed, the accuracy of the action detection module is influenced by several parameters: the threshold, the sampling rate, and the time step.
To illustrate the effects of these parameters, we collect some instances and plot the power difference between two consecutive windows 
in Fig.~\ref{fig:energy_instance}.
It can be seen that the power difference in the second phase is distinctly higher than that in the first phase, confirming the effectiveness of the action detection module. Additionally, the power difference gap between the two phases becomes larger with the sampling rate or the time step, which further enhances the performance of action detection.
The above analyses indicate that it is reasonable and necessary to optimize above parameters
to guarantee the performance of the action detection module. 
However, the challenge lies in the fact that the expression for calculating power difference outlined in Section~\ref{sec:action_detection} does not explicitly describe the mathematical relationship among the sampling rate, threshold, time step, miss rate, and false positive rate. In the following, we aim to address this challenge.

We start with analyzing the power difference in~\eqref{eq:delta_p}. According to~\cite{he2023integrated}, when a time window is filled with the variation caused by the $i$-th action type, the power of the high-frequency components, denoted by $P^{\mathrm{c}}_{k,i}$, follows a normal distribution, and the mean and variance are
\begin{align}
 \mu^{\mathrm{P}}_{k,i} &=  \lambda_i + \frac{r_i}{T^{\mathrm{s}} F^{\mathrm{s}}_k},\\ 
  (\sigma^{\mathrm{P}}_{k,i})^2 &= 4 \frac{\sigma^2_{\mathrm{c}}}{T^{\mathrm{s}}F^{\mathrm{s}}_k}\lambda_i  + 2 \frac{\sigma^2_{\mathrm{c}} }{\left( T^{\mathrm{s}} F^{\mathrm{s}}_k\right)^2}r_i, \label{eq:mu and sigma}
\end{align}
respectively, where $\lambda_i$, $r_i$, and $\sigma^2_{\mathrm{c}}$
are parameters that can be acquired via fitting on collected CSI. We aim to extend this model for $\Delta P_k[l]$. Recall that $\Delta P_k[l]$ is the power difference between the current and previous time windows. Meanwhile, there is overlap between two consecutive time windows and the duration of the overlap is $T^{\mathrm{s}}\!-\! \tau_k$, as shown in Fig.~\ref{fig:sec-phase}. Thus, the power can be regarded as a linear combination of the power of the overlap and the current time step, as
\begin{equation}
    P_k[l]=\frac{T^{\mathrm{s}} - \tau_k}{T^{\mathrm{s}}}P^{\mathrm{c}}_{k,1}+\frac{\tau_k}{T^{\mathrm{s}}}P^{\mathrm{c}}_{k,i}.
\end{equation}
Since the sensing begins with the first phase (i.e., the static state), the overlap is in the first phase, and the current time step would be in the first phase or the second phase.
As the power overlap contributes the same for the two time windows, the power difference can be regarded as the difference between the current and previous time steps. Consequently, when both time steps are in the first phase, the mean and variance of $\Delta P_k[l]$ are 
\begin{align} 
   \mu_{k,1}^{\Delta}&= \frac{\tau_k}{T^{\mathrm{s}}}\mu_{k,1}^{\mathrm{P}}-\frac{\tau_k}{T^{\mathrm{s}}}\mu_{k,1}^{\mathrm{P}}=0, \label{eq:mu_static}\\
   (\sigma^{\Delta}_{k,1})^2 &=\left(\frac{\tau_k}{T^{\mathrm{s}}}\sigma^{\mathrm{P}}_{k,1}\right)^2+ \left(\frac{\tau_k}{T^{\mathrm{s}}}\sigma^{\mathrm{P}}_{k,1}\right)^2=2\left(\frac{\tau_k}{T^{\mathrm{s}}}\sigma^{\mathrm{P}}_{k,1}\right)^2,     
\end{align}
respectively. 

As the time window moves forward, the current time step begins to contain variation caused by the $i$-th action type, and the variation's duration is $t_k$ with $t_k\in[0,\tau_k]$. However, there is no variation in the previous time step.
We aim to detect this case to obtain the onset of the 
action. In this case, the mean and variance of $\Delta P_k[l]$ can be expressed by
\begin{align}
    \mu_{k,i}^{\Delta}(t_k)&= \frac{t_k}{T^{\mathrm{s}}}\mu_{k,i}^{\mathrm{P}} + \frac{\tau_k-t_k}{T^{\mathrm{s}}}\mu_{k,1}^{\mathrm{P}} - \frac{\tau_k}{T^{\mathrm{s}}}\mu_{k,1}^{\mathrm{P}} \nonumber\\
    &= \frac{t_k}{T^{\mathrm{s}}} \left( \mu_{k,i}^{\mathrm{P}} - \mu_{k,1}^{\mathrm{P}}\right), \\
    (\sigma^{\Delta}_{k,i})^2(t_k)&= \left(\frac{t_k}{T^{\mathrm{s}}}\sigma^{\mathrm{P}}_{k,i}\right)^2\! +\! \left(\frac{\tau_k-t_k}{T^{\mathrm{s}}}\sigma^{\mathrm{P}}_{k,1}\right)^2 \!+\! \left(\frac{\tau_k}{T^{\mathrm{s}}}\sigma^{\mathrm{P}}_{k,1}\right)^2\nonumber\\
    &= \left(\frac{t_k}{T^{\mathrm{s}}}\sigma^{\mathrm{P}}_{k,i}\right)^2\! +\! \frac{2\tau_k^2 + t_k^2 - 2\tau_k t_k}{(T^{\mathrm{s}})^2}\left(\sigma^{\mathrm{P}}_{k,1}\right)^2, 
\end{align}
respectively, with $i\ge 2$.
Since the onset of an action follows a Poisson process, $t_k$ follows the uniform distribution and the expectation of $\mu_{k,i}^{\Delta}$ and $(\sigma^{\Delta}_{k,i})^2$ can be given by
\begin{align} 
  \mu_{k,i}^{\Delta}\!&= \frac{1}{\tau_k}\int_{0 }^{\tau_k}\mu_{k,i}^{\Delta}(t_k) \mathrm{d} t_k \nonumber\\
 &=\tau_k\left(\frac{\lambda_i-\lambda_1}{2T^{\mathrm{s}}}+\frac{r_i-r_1}{ 2(T^{\mathrm{s}})^2F^{\mathrm{s}}_k}\right), \label{eq:mu_active_mean} \\
  (\sigma^{\Delta}_{k,i})^2\!&=\frac{1}{\tau_k}\int_{0 }^{\tau_k}(\sigma^{\Delta}_{k,i})^2(t_k)\mathrm{d} t_k \nonumber \\
  &=\tau_k^2\!\left(
    \frac{4\sigma^2_{\mathrm{c}}(\lambda_i\!+\!4\lambda_1)}{3(T^{\mathrm{s}})^3F^{\mathrm{s}}_k}\! 
    +\! \frac{2\sigma^2_{\mathrm{c}}(r_i+4r_1)}{3(T^{\mathrm{s}})^4 (F^{\mathrm{s}}_k )^2}
  \right),
\end{align}
respectively. Additionally, instances of the same action type would not be identical. 
Thus, we introduce a constant $\sigma^2_{\mathrm{d},i}$ 
to represent the deviation, and
the revised variance is 

\begin{align}
\!\!\!\!(\sigma^{\Delta}_{k,i})^2 \!=\!\left\{
\begin{array}{ll}
    \!\!\!\tau_k^2\left( \frac{8\sigma^2_{\mathrm{c}}}{(T^{\mathrm{s}})^3F^{\mathrm{s}}_k}\lambda_1 \! +  \!\frac{4\sigma^2_{\mathrm{c}} }{(T^{\mathrm{s}})^4 (F^{\mathrm{s}}_k )^2}r_1\right)\!+\!\sigma^2_{\mathrm{d},1}, & \!\!\!\!i\!=\!1,\!\\
    \!\!\!\tau_k^2\!\left(
    \frac{4\sigma^2_{\mathrm{c}}(\lambda_i\!+\!4\lambda_1)}{3(T^{\mathrm{s}})^3F^{\mathrm{s}}_k}\! 
   \! +\! \frac{2\sigma^2_{\mathrm{c}}(r_i+4r_1)}{3(T^{\mathrm{s}})^4 (F^{\mathrm{s}}_k )^2}
  \right)\!+\!\sigma^2_{\mathrm{d},i}, & \!\!\!\!i\!\ge\!2.\!
\end{array}\right. \label{eq:sigma_active_mean}
\end{align}

Until now, we can mathematically formulate the relationship among the sampling rate, time step, threshold, miss rate, and false positive rate, as shown in the following proposition.
\begin{prop}\label{prop:rate}
 The miss rate of the $i$-th action type for device $k$ is given by
\begin{equation} \label{eq:miss}
    p_{k,i}^{\mathrm{o}}=Q\left(\dfrac{\mu^{\Delta}_{k,i}-\eta_k}{\sigma^{\Delta}_{k,i}}\right),
\end{equation}
and the false positive rate for device $k$ is
\begin{equation}\label{eq:false}
    p_k^{\mathrm{l}}=Q\left(\dfrac{\eta_k}{\sigma^{\Delta}_{k,1}}\right),
\end{equation}
where $Q(\cdot)$ is the Q-function.
\end{prop}

With Proposition~\ref{prop:rate}, we can find that $p_k^{\mathrm{l}}$ decreases but $p_{k,i}^{\mathrm{o}}$ increases with the threshold when $\eta_k\! \in \![0,\min_{i\in \{2,\cdots, I\}}{\!\mu^{\Delta}_{k,i}}]$, 
which matches the discussion in Section~\ref{sec:action_detection}. Meanwhile, both the mean and variance of $\Delta P_k[l]$ decrease with the sampling rate, which aligns with Fig.~\ref{fig:energy_instance}. 
\begin{figure}[t]
	\setlength{\abovecaptionskip}{8pt} 	
	\centering
	\includegraphics[width=0.75\linewidth]{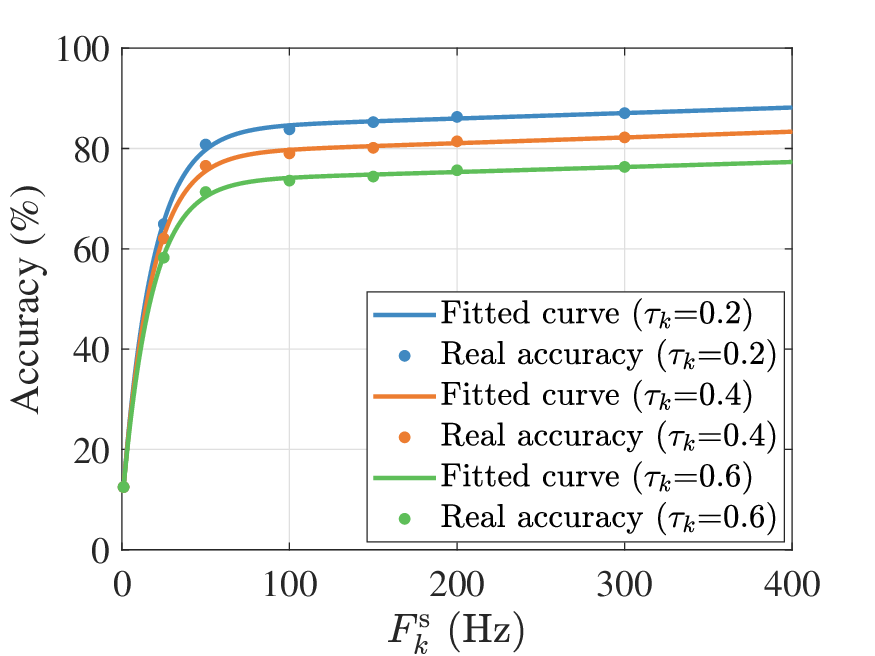}
	\vspace{-2ex}
	\caption{The recognition accuracy with different sampling rates and time steps. The test setting and dataset are presented in Section~\ref{sec:test}.}
	\label{fig:acc_F_initial}
    \vspace{-2ex}
\end{figure}

\subsubsection{Edge Recognition Module}
According to~\cite{liu2023rethink}, the accuracy of the CNN-based edge recognition module increases monotonically with the sampling rate.
Moreover,
the analysis in Section~\ref{sec:action_detection}
indicates that the duration of the signal of interest within the window that inputs to the edge CNN module is randomly distributed over the interval $[T^{\mathrm{s}}\!-\!\tau_k/2,T^{\mathrm{s}}]$.
This implies that a larger time step leads to less information from the sensing targets on average, resulting in a negative correlation between the accuracy and time step. 
Therefore, the accuracy of the CNN module can be modeled as a function that monotonically increases with the sampling rate and monotonically decreases with the time step, as $\alpha_k(F^{\mathrm{s}}_k,\tau_k)$. To confirm this, we collect a sensing dataset and adopt a four-layer CNN to realize the action recognition.
The relationship between the recognition accuracy and sampling rate under different time steps is shown in Fig.~\ref{fig:acc_F_initial}.
We can observe that the results are consistent with the aforementioned analysis. When the false positive rate and miss rate are almost zeros, the average sensing accuracy for the device $k$ is
\begin{equation}
    A_k = q_{k,1}+ \sum_{i=2}^{I}q_{k,i}\alpha_k(F^{\mathrm{s}}_k,\tau_k),
\end{equation}
where $q_{k,i}$ represents the probability of the $i$-th action type for device $k$ and
$\sum_{i=1}^{I}q_{k,i}=1$.

\subsection{Power Consumption}
Since the BS is powered externally, we only focus on the power of IoT devices. It consists of three parts:
sensing power for transmitting sensing signals, 
local computation power, and wireless transmission power. 

For the sensing power, let $E^{\mathrm{s}}_k$ denote the energy consumption of device $k$ for each sensing transmission, and the average sensing power is $P^{\mathrm{sen}}_k = E^{\mathrm{s}}_kF^{\mathrm{s}}_k$. 
For the local computation, the main computation cost is caused by the FFT operation. Each IoT device needs to process a CSI vector with the size being $(T^{\mathrm{s}}F^{\mathrm{s}}_k)\times 1$ after each time step. Let $E^\mathrm{c}_k$ represent the energy consumption per element for device $k$, and then the computation power can be expressed as  
\begin{equation}
P_k^{\mathrm{comp}}=\frac{E^\mathrm{c}_kT^{\mathrm{s}}F_k^{\mathrm{s}}}{\tau_k}.
\end{equation}

For wireless transmission, the IoT device does not need to send each CSI matrix to the edge server. Instead, when both the miss rate and false positive rate approach zero, the probability of transmission is $(1-q_{k,1})$. Moreover, during each transmission, the time-division manner
is adopted for realizing ISAC, and the proportion allocated to communication is $(1-t^{\mathrm{s}} F_k^{\mathrm{s}})$. Thus, the average communication power for the $k$-th device is
\begin{equation}
    P_k^{\mathrm{trans}}=(1-t^{\mathrm{s}} F_k^{\mathrm{s}})P_{k}^{\mathrm{t}}(1-q_{k,1}).
\end{equation}

Based on the above analysis, the overall power for the $k$-th device is 
\begin{equation}
    P_k^{\mathrm{over}} =  P^{\mathrm{sen}}_k + P_k^{\mathrm{comp}}+P_k^{\mathrm{trans}}. 
\end{equation}

\subsection{Sensing Delay}
The delay of the sensing task also contains three parts: computation delay of the action detection module, wireless transmission delay, and edge recognition delay. The delay of the action detection module mainly comes from the FFT operation, which can be expressed as
\begin{equation}
T_k^{\mathrm{comp}}=\frac{T^{\mathrm{s}}F_k^{\mathrm{s}}C^\mathrm{L}}{f^\mathrm{L}_k},
\end{equation}
where $C^\mathrm{L}$ represents the computational intensity of FFT operation (in CPU cycles/element), and $f^\mathrm{L}_k$ (in CPU cycles/s, i.e., Hz) denotes the computation resource of device $k$.

The wireless transmission delay increases with the size of the sensing data and decreases with the communication resource. Let $V^\mathrm{L}$ (in bits/element) denote the data size per element, and then the corresponding delay for device $k$ can be calculated as 
\begin{equation}    T_k^{\mathrm{trans}}=\frac{NT^{\mathrm{s}}F_k^{\mathrm{s}}V^\mathrm{L}(1-q_{k,1})}{(1-t^{\mathrm{s}} F_k^{\mathrm{s}})R_k},
\end{equation}
where $(1-q_{k,1})$ represents the probability
of transmission. 

According to~\cite{he2020optimizing}, the computational complexity of convolutional layers is proportional to the input data size, i.e., $N \times T^{\mathrm{s}}F_k^{\mathrm{s}}$. Then, the delay for the CNN module corresponding to device $k$ can be expressed as
\begin{equation} \label{eq:com_cnn}
T_k^{\mathrm{CNN}}=\frac{NT^{\mathrm{s}}F_k^{\mathrm{s}}C^\mathrm{e}(1-q_{k,1})}{f_k^\mathrm{e}},
\end{equation}
where $f_k^\mathrm{e}$ denotes the computation resource allocated to device $k$ and $C^\mathrm{e}$ represents the computation cost per element.

\subsection{Problem Formulation}
In this paper, we aim to maximize the sensing performance of all IoT devices. To this end, we take the average sensing accuracy of all IoT devices as the objective function, which represents the overall sensing performance of the considered system. Therefore, the sensing accuracy maximization problem can be formulated as
\begin{subequations} \label{pb_o}
\begin{eqnarray}
	\!\!&\max\limits_{\left\{  F^{\mathrm{s}}_k, \eta_k, f_k^\mathrm{e}, \tau_k  \right\}} & \!\!A \!=\! \frac{1}{K} \!\sum_{k=1}^{K} \!\left(\!q_{k,1}\!+\!\sum_{i=2}^{I}q_{k,i}\alpha_k(F^{\mathrm{s}}_k,\tau_k)\!\right)\!, ~~~~~\\
	&\text{s.t.} &\!\! p_{k,i}^{\mathrm{o}} \leq p^{\min}, ~i=2,3,\cdots,I,   \label{stt1}\\ 
    &   & \!\!p_k^{\mathrm{l}} \leq p^{\min}, \label{stt2}\\
    &   &\!\! T_k^{\mathrm{comp}}\leq \tau_k ,   \label{stt3}\\
	&	&\!\! T_k^{\mathrm{trans}}+T_k^{\mathrm{CNN}} \leq T^{\max} , \label{stt4}\\
	&	&\!\! P_k^{\mathrm{over}} \leq P_k^{\max} ,  \label{stt5}\\
	&	&\!\! \sum_{k=1}^{K}f_k^\mathrm{e} \leq f^{\mathrm{e}},	\label{stt6}\\
        &	&\!\! F^{\mathrm{s}}_k \in \mathcal{Z}^+,   \eta_k,f_k^\mathrm{e}, \tau_k \ge 0,	\label{stt7}
\end{eqnarray}
\end{subequations}
where $P^{\max}_k$ is the power upper limit, $T^{\max}$ is the delay upper limit, $f^{\mathrm{e}}$ is the total computation resource at the edge server, and $\mathcal{Z}^+$ denotes the set of positive integers. To ensure the performance of the proposed action detection module, constraints~(\ref{stt1}) and~(\ref{stt2}) require the miss rate and false positive rate to be under their requirements (i.e., $p^{\min}$), and $p^{\min}$ is set to a small value (e.g., 2~\!\% in Section~\ref{subsec:test}) to mitigate misdetection.
Constraint~(\ref{stt3}) limits the computation time for the action detection module to avoid jam. Constraints~(\ref{stt4}) and~(\ref{stt5}) limit the delay and power of the sensing task, respectively. Moreover, constraint~(\ref{stt6}) represents the edge computation resource constraint.

\section{Sensing Accuracy Maximization}\label{sec:solve}

In this section, we first decompose problem~(\ref{pb_o}) into two subproblems and then propose a sensing accuracy maximization algorithm by solving them.

\subsection{Problem Decomposition}
The problem~(\ref{pb_o}) is intractable due to the highly coupled and non-convex objective function and constraints. To solve it, we opt to decompose it.
Because of the local-edge interactive nature in the considered ISCC system, it is reasonable to consider decomposing the optimization problem into two types of subproblems with the help of ADMM method~\cite{wang2019global}: one for each IoT device and the other for the edge server. They can be solved separately at the local and edge levels. 
Note that $f_k^\mathrm{e}$ influences both the sensing delay for the IoT device and the resource allocation at the edge server.  
Therefore, before performing the ADMM method, we first introduce the auxiliary variable $\hat{f}^\mathrm{e}_{k}$, which meets the following linearly coupled equality (LCE) constraints:
\begin{equation} \label{stt8}
    \hat{f}^\mathrm{e}_{k} = f^{\mathrm{e}}_k,~k=1,\cdots, K. 
\end{equation}
Then, constraint~(\ref{stt4}) can be rewritten as
\begin{equation} \label{stt9}   \frac{NT^{\mathrm{s}}F_k^{\mathrm{s}}V^\mathrm{L}(1-q_{k,1})}{(1-t^{\mathrm{s}} F_k^{\mathrm{s}})R_k}+\frac{NT^{\mathrm{s}}F_k^{\mathrm{s}}C^\mathrm{e}(1-q_{k,1})}{ \hat{f}^\mathrm{e}_{k}} \leq T^{\max}. 
\end{equation}
Consequently, problem~(\ref{pb_o}) can be equivalently converted into
\begin{eqnarray}  \label{pb_o1}
	\!\!\!&\max\limits_{\left\{  F_k^{\mathrm{s}}, \eta_k, f_k^\mathrm{e}, \tau_k ,\hat{f}^\mathrm{e}_{k} \right\}} & \!\!\!\!A \!=\! \frac{1}{K} \!\sum_{k=1}^{K} \!\left(\!q_{k,1}\!+\!\sum_{i=2}^{I}q_{k,i}\alpha_k(F^{\mathrm{s}}_k,\tau_k)\!\right)\!, ~~~~~\\
	\!\!\!&\text{s.t.} &  \!\!\!\!\text{(\ref{stt1})--(\ref{stt3}),~(\ref{stt5})--(\ref{stt7}),~(\ref{stt8}), and~(\ref{stt9})}.\nonumber
\end{eqnarray}
After dualizing and penalizing the LCE constraints into the objective function with Lagrange multipliers, denoted by $\{\beta_1,\cdots,\beta_K\}$, the augmented Lagrangian (AL) problem can be derived as
\begin{eqnarray} \label{pb_o2}
	&\max\limits_{\left\{  \mathbb{X}_k, \mathbb{F}\right\}} &  \mathcal{L}(\mathbb{X}_1,\cdots,\mathbb{X}_K,\mathbb{F},\beta_1,\cdots,\beta_K)  \nonumber\\
 & &=\frac{1}{K} \!\sum_{k=1}^{K} \!\left(\!q_{k,1}\!+\!\sum_{i=2}^{I}q_{k,i}\alpha_k(F^{\mathrm{s}}_k,\tau_k)\!\right)\! \nonumber\\
 &&- \frac{1}{K}\sum_{k=1}^K\frac{\rho}{2} ||   f^\mathrm{e}_k-\hat{f}^\mathrm{e}_{k} + \frac{1}{\rho} \beta_k||^2, \\ 
	&\text{s.t.} &  \text{(\ref{stt1})--(\ref{stt3}),~(\ref{stt5})--(\ref{stt7}), and~(\ref{stt9})},   \nonumber
\end{eqnarray}
where $\mathbb{X}_k=\{F^{\mathrm{s}}_k, \eta_k, \hat f_k^\mathrm{e}, \tau_k \}$, $k=1, 2,\cdots,K$, $\mathbb{F}=\{f^\mathrm{e}_1,f^\mathrm{e}_2,\cdots,f^\mathrm{e}_K\}$, and $\rho$ is the penalty parameter. 
Then, the optimization variables can be divided into $K+1$ blocks: $\mathbb{X}_1$, $\cdots$, $\mathbb{X}_K$, and $\mathbb{F}$.
The first $K$ blocks are optimized to maximize the sensing accuracy of each IoT device, separately. 
The block $\mathbb{F}$ is optimized for resource allocation at the edge server. 
With fixing other optimization variables, the accuracy maximization subproblem for device $k$ is given by
\begin{eqnarray} \label{pb_o3}
        &\max\limits_{ \mathbb{X}_k } &  \!\!q_{k,1}\!+\!\sum_{i=2}^{I}q_{k,i}\alpha_k(F^{\mathrm{s}}_k,\tau_k)\!-\! \frac{\rho}{2} ||   f^\mathrm{e}_k\!-\!\hat{f}^\mathrm{e}_{k} \!+\! \frac{1}{\rho} \beta_k||^2, ~~~~~\\
        &\text{s.t.} & \!\!\text{(\ref{stt1})--(\ref{stt3}),~(\ref{stt5}),~(\ref{stt7}), and~(\ref{stt9})}. \nonumber
\end{eqnarray}
Similarly, the subproblem for resource allocation with the block $\mathbb{F}$ is given by
\begin{eqnarray} \label{pb_o4}
        &\min\limits_{\mathbb{F}} & 
 \frac{1}{K} \sum_{k=1}^K\frac{\rho}{2} || f^\mathrm{e}_k-\hat{f}^\mathrm{e}_{k} + \frac{1}{\rho} \beta_k||^2,  \\
        &\text{s.t.} & \text{(\ref{stt6}) and~(\ref{stt7})}.\nonumber
\end{eqnarray}
In each iteration, we first need to solve the above two types of subproblems, 
whose detailed solution will be introduced in the next subsection.
After that, we need to update the dual variables in the $i$-th iteration, as
\begin{equation} \label{eq:lar}
    \beta_k^{(i+1)} = \beta_k^{(i)} + \rho (f^{\mathrm{e}}_k - \hat{f}^{\mathrm{e}}_k), ~ k=1, 2, 
    \cdots,K. 
\end{equation}

\subsection{Sensing Accuracy Maximization for Each Device}
First of all, we focus on problem~(\ref{pb_o3}). One can clearly observe that the threshold $\eta_k$ in the action detection module does not appear at the objective function, and thus we first focus on it to reduce the complexity of problem~(\ref{pb_o3}). The selection of the threshold can be transformed into the existence of a threshold that simultaneously satisfies constraints~(\ref{stt1}) and~(\ref{stt2}). 
Recall that $p_k^{\mathrm{l}}$ decreases but $p_{i,k}^{\mathrm{o}}$ increases with the threshold. As we can imagine, given the sampling rate $F_k^{\mathrm{s}}$ and time step $\tau_k$, the curves of $p_k^{\mathrm{l}}$ and $p_{i,k}^{\mathrm{o}}$ with the threshold will intersect at a unique point, representing the minimum achievable values for both $p_k^{\mathrm{l}}$ and $p_{i,k}^{\mathrm{o}}$, denoted as $p^\mathrm{c}_{k}$.
For $p^\mathrm{c}_{k}$, we have the following theorem.
\begin{thm}\label{thm:p}
    Under given $F_k^{\mathrm{s}}$, $p^\mathrm{c}_{k}$ decreases with $\tau_k$.
    \begin{IEEEproof}
	Please refer to Appendix~\ref{proof:p}.
\end{IEEEproof}
\end{thm}

Theorem~\ref{thm:p} is reasonable, as a larger time step means more collected information,
thus enhancing the performance of the action detection module. Furthermore, with Theorem~\ref{thm:p}, we can observe that the time step must exceed a certain value for a given sampling rate to satisfy constraints~(\ref{stt1}) and~(\ref{stt2}). Specifically, the following constraint should be met
\begin{equation}\label{trans1}
    p_{k,i}^{\mathrm{o}}=p_k^{\mathrm{l}}\le p^{\min}.
\end{equation}
After selecting the threshold for satisfying ``='' in the above constraint, both $p_{k,i}^{\mathrm{o}}$ and $p_k^{\mathrm{l}}$ are the functions of $\tau_k$ and $F^{\mathrm{s}}_k$. Now, we can use a function of $F_k^{\mathrm{s}}$, denoted by $u(F^{\mathrm{s}}_k)$, to represent the required $\tau_k$ for satisfying $p_k^{\mathrm{l}}=p^{\min}$. Then, with Theorem~\ref{thm:p}, to ensure that both $p_{k,i}^{\mathrm{o}}$ and $p_k^{\mathrm{l}}$ are no more than $p^{\min}$, we need to satisfy the following constraint:
 \begin{equation}\label{st:rate}
   u(F^{\mathrm{s}}_k) \leq \tau_k.
 \end{equation}
Note that
$u(F^{\mathrm{s}}_k)$ can be established after collecting sensing dataset, 
and it can be directly utilized in the proposed algorithm with a constant time complexity.

Similarly, constraint~(\ref{stt5}) can be translated into a relationship between the sampling rate and time step, as 
\begin{equation}\label{st:power}
    \tau_k \geq \frac{E^\mathrm{c}_kT^{\mathrm{s}}F_k^{\mathrm{s}}}{P_k^{\max}\!\!-\!\!(1\!\!-\!\!t^{\mathrm{s}} F_k^{\mathrm{s}})P_k^{\mathrm{t}}(1\!\!-\!\!q_{k,1})\!\!-\!\!E^{\mathrm{s}}_kF^{\mathrm{s}}_k}=v(F^{\mathrm{s}}_k).
\end{equation}

To maximize the sensing accuracy, the time step should be taken as the minimum value that simultaneously satisfies constraints~(\ref{st:rate}),~(\ref{st:power}), and (\ref{stt3}), as
 \begin{equation}\label{st:tau}
     \tau^\star_k=\max \{ u(F^{\mathrm{s}}_k), v(F^{\mathrm{s}}_k), \frac{T^{\mathrm{s}}F_k^{\mathrm{s}}C^\mathrm{L}}{f^{\mathrm{l}}_k} \}. 
 \end{equation} 
 
Moreover, we can observe that the constraint~(\ref{stt9}) primarily describes the relationship between the sampling rate and $\hat{f}^{\mathrm{e}}_{k}$, which can be transformed
into
 \begin{equation}\label{st:f}
     \hat f_k^\mathrm{e} \geq \frac{NT^{\mathrm{s}}F_k^{\mathrm{s}}C^\mathrm{e}(1-q_{k,1})}{T^{\max}-\dfrac{NT^{\mathrm{s}}F_k^{\mathrm{s}}V^\mathrm{L}(1-q_{k,1})}{(1-t^{\mathrm{s}} F_k^{\mathrm{s}})R_k}}=\gamma(F^{\mathrm{s}}_k). 
 \end{equation}
Thus, 
problem~(\ref{pb_o3}) can be reformulated as 
 \begin{eqnarray} \label{pb_o5}
        &\!\!\!\max\limits_{\{F^{\mathrm{s}}_k, \hat{f}_k^{\mathrm{e}}, \tau_k \} } &  \!\!\!\!\!\!q_{k,1}\!+\!\sum_{i=2}^{I}q_{k,i}\alpha_k(F^{\mathrm{s}}_k,\tau_k)\!-\! \frac{\rho}{2} || f^\mathrm{e}_k\!\!-\!\hat{f}^\mathrm{e}_{k} \!+\! \frac{1}{\rho} \beta_k||^2,~~~~~\\
        &\text{s.t.} &  \!\!\!\!\!\ \text{(\ref{st:tau})~and~(\ref{st:f})}.  \nonumber
\end{eqnarray}

To solve problem~(\ref{pb_o5}), we first focus on $\hat{f}_k^{\mathrm{e}}$ under given $F^{\mathrm{s}}_k$, and the objective function is a quadratic function with $\hat{f}^\mathrm{e}_{k}$. Thus, there are two cases.
\begin{itemize}
    \item When $\gamma(F^{\mathrm{s}}_k) \leq {f}^\mathrm{e}_{k}+\frac{1}{\rho} \beta_k$, $\hat{f}^\mathrm{e}_{k}$ can be directly set to ${f}^\mathrm{e}_{k}+\frac{1}{\rho} \beta_k$, and the value of the penalty function is zero.
    \item When $\gamma(F^{\mathrm{s}}_k) > {f}^\mathrm{e}_{k}+\frac{1}{\rho} \beta_k$, the objective function decreases with $\gamma(F^{\mathrm{s}}_k)$. The maximum accuracy is achieved when $\hat{f}^\mathrm{e}_{k}=\gamma(F^{\mathrm{s}}_k) $.
\end{itemize}
To conclude, the optimal 
$\hat{f}^\mathrm{e}_{k}$ can be denoted as
\begin{equation} \label{st:f_fin}
    \hat f_k^{\mathrm{e},\star}=\max \{ {f}^\mathrm{e}_{k}+\frac{1}{\rho} \beta_k, \gamma(F^{\mathrm{s}}_k) \}. 
\end{equation}

Based on the above analysis,  both $\tau_k$ and $\hat f_k^\mathrm{e}$ can be regarded as functions of the sampling rate $F^{\mathrm{s}}_k$. Thus, problem~(\ref{pb_o3}) can be transformed into an unconstrained optimization problem with respect to the sampling rate. Specifically, when the sampling rate is $F_k^{\mathrm{s}}$, the accuracy can be calculated as
\begin{equation}\label{eq:acc}
A_k^{{\star}}=q_{k,1}+\sum_{i=2}^{I}q_{k,i}\alpha_k(F^{\mathrm{s}}_k,\tau_k^\star) - \frac{\rho}{2} ||   f^{\mathrm{e}}_k-\hat{f}_k^{\mathrm{e},\star} + \frac{1}{\rho} \beta_k||^2.
\end{equation} 
Since $F^{\mathrm{s}}_k$ is a positive integer with an upper limit, the maximum accuracy for device $k$ can be obtained via the exhaustive search algorithm on $F^{\mathrm{s}}_k$, 
and the algorithm is summarized in Algorithm~\ref{alg:single}.
To narrow the search space, we give an upper bound of $F_k^{\mathrm{s}}$ under the constraint~(\ref{st:f}) on transmission delay, as
\begin{equation}
    F^{\mathrm{s}, \max}_k=\frac{1}{t^{\mathrm{s}} +\dfrac{NT^{\mathrm{s}}V^\mathrm{L}(1-q_{k,1})}{T^{\max}R_k}}.
\end{equation}

\begin{algorithm}[t]
		\caption{Algorithm for Solving Problem~({\ref{pb_o3}}).}
		\label{alg:single}
		\DontPrintSemicolon            
		\textbf{Initialize:} the sensing accuracy $A_k=0$;\;
            \For{$F^{\mathrm{s}}_k \in \left\{1,\cdots,F^{\mathrm{s}, \max}_k\right\}$}{
            Obtain $\tau^{\star}_k$ with~(\ref{st:tau});\;
            Obtain $\hat f_k^{\mathrm{e},{\star}}$ with~(\ref{st:f_fin});\;
            Calculate the accuracy $A_k^{\star}$ with~(\ref{eq:acc});\;
            \If{$ A_k<A_k^{{\star}} $}
            {$ A_k=A_k^{{\star}}$, $\hat f_k^{\mathrm{e}}=\hat f_k^{\mathrm{e},{\star}}$;}
            }		
\end{algorithm} 

\subsection{Edge Resource Allocation}

It can be easily confirmed that the problem~(\ref{pb_o4}) is convex, since the objective function is quadratic and the constraints are linear.
Then, we can employ the Lagrangian method to obtain the optimal solution to problem~(\ref{pb_o4}),  
and the partial Lagrangian function
can be expressed as 
\begin{align} \label{pb_o6}
        \mathcal{L}^{\mathrm{e}}=
 \frac{1}{K} \sum_{k=1}^K\frac{\rho}{2} || f^\mathrm{e}_k-\hat{f}^\mathrm{e}_{k} + \frac{1}{\rho} \beta_k||^2 + \mu(\sum_{k=1}^{K}f_k^\mathrm{e} - f^{\mathrm{e}} ),
\end{align}
where $\mu\ge0$ is the Lagrange multiplier associated with constraint~(\ref{stt6}). By applying the Karush-Kuhn-Tucker (KKT) conditions, we can obtain the optimal solution to problem~(\ref{pb_o6}) in the following theorem. 
\begin{thm} \label{thm:resource}
   Let $f_k^{\mathrm{e},\star}$ denote the optimal solution to problem~(\ref{pb_o6}), and it can be given by
\begin{eqnarray}\label{eq:f_opti}
    f_k^{\mathrm{e},\star} =    
         \max \{\hat{f}^\mathrm{e}_{k}\!-\!\dfrac{\beta_k}{\rho}\!-\!\dfrac{K}{\rho}\mu^{\star},0 \},
\end{eqnarray}
where $\mu^{\star}$ satisfies $\sum_kf_k^{\mathrm{e},\star} \leq f^{\mathrm{e}} $. 
\begin{IEEEproof}
	Please refer to Appendix~\ref{proof:resource}.
\end{IEEEproof}
\end{thm}

Theorem~\ref{thm:resource} gives the optimal resource allocation strategy. There are two cases. When the edge computation resource is sufficient (i.e., $\mu^{\star}=0$),
each device can be allocated the required resource estimated locally. When the edge resource is insufficient, 
the resource allocated to each device is negatively related to $\mu^{\star}$. This is reasonable since $\mu^{\star}$ represents the degree of violation of constraint~(\ref{stt6}) under $\hat{f}^{\mathrm{e}}_k$.

With Theorem~\ref{thm:resource},  Algorithm~\ref{alg:KKT} is developed to find $f_k^{\mathrm{e},\star}$.
We first set $\mu^{\star} = 0$ and check whether the constraint $\sum_kf_k^{\mathrm{e},\star}\le f^{\mathrm{e}}$ is satisfied. If satisfied,
$f_k^{\mathrm{e},\star}$ is directly output as the optimal solution. Otherwise,
the bisection search algorithm is employed until $|\sum_{k=1}^Kf_k^{\mathrm{e},\star} -f^{\mathrm{e}}|<\epsilon$ and the optimal solution can be obtained. To accelerate the search, we give the upper and lower bounds for $\mu^{\star}$ as
\begin{align}
\left \{
    \begin{array}{ll}
        \mu^{\star}\geq \mu_{\min}=0,\\
		  \mu^{\star}\leq \mu_{\max}=\frac{\rho}{K}\max_k\{\hat{f}^\mathrm{e}_{k}-\frac{\beta_k}{\rho} \}.
\end{array} 
\right.
\end{align}
The upper bound can be easily obtained by setting $f_k^{\mathrm{e}}$ to zero.

\begin{algorithm}[t]
    \caption{Edge Resource Allocation Algorithm.}
    \label{alg:KKT}
    \DontPrintSemicolon  
    Set the maximal error tolerance $\epsilon>0$, $\mu^{\star}=0$;\;
    Calculate $f_k^{\mathrm{e},\star}=\max \{\hat f_k^\mathrm{e}-\dfrac{\beta_k}{\rho},0\}$;\;
        \eIf{$\sum_{k=1}^K f_k^{\mathrm{e},\star} \leq f^{\mathrm{e}}$}
            {\textbf{Output:} $f_k^{\mathrm{e},\star}$;}
        {Set $\mu_l=\mu_{\min}$, $\mu_u=\mu_{\max}$;\;
        \Repeat{$|\sum_{k=1}^Kf_k^{\mathrm{e},\star} -f^{\mathrm{e}}|<\epsilon$;}{Let $\mu=(\mu_{l}+\mu_{u})/2$;\;
        Calculate $f_k^{\mathrm{e},\star}=\max \{\hat{f}^\mathrm{e}_{k}\!-\!\dfrac{\beta_k}{\rho}\!-\!\dfrac{K}{\rho}\mu,0 \}$;\;
    \eIf{$\sum_{k=1}^K f_k^{\mathrm{e},\star} > f^{\mathrm{e}}$}{
    $\mu_l = \mu$;\;  
    }{
        $\mu_u = \mu$;\;
    }
}
\textbf{Output:} $\mu^{\star}=\mu$ and $f_k^{\mathrm{e},\star}$.}
\end{algorithm} 

\subsection{Overall Algorithm}
{\begin{algorithm}[t]
		\caption{ADMM-based Distributed Algorithm for Problem~({\ref{pb_o}}).}
		\label{alg:overall}
		\DontPrintSemicolon            
		\textbf{Initialize:} $\beta_k$, $f_k^\mathrm{e}$, $\rho$, the maximal primal residual $\varepsilon>0$, the current iteration number $ i = 0$, and the maximum iteration number $I_{\max}$;\;
        \Repeat{$\|f_k^\mathrm{e,\star} - \hat f_k^\mathrm{e}\|_2 \leq \varepsilon$, $\forall k$ \textnormal{or} $i>I_{\max}$;}
        {
         \% Accuracy maximization at the device;\;
         \For{$k=1,2,\cdots, K$}{
         Obtain $A_k$ and $\hat f_k^{\mathrm{e}}$ using Algorithm~\ref{alg:single};}
			\% Resource allocation at the edge server;\;
			Obtain $ f_k^\mathrm{e,\star}$ with Theorem~\ref{thm:resource};\;
                Update $\beta_k$ according to~(\ref{eq:lar}); \;
            $i = i + 1$;\;
		}
      \textbf{Output:} the overall accuracy $A=\frac{1}{K} \sum_{k=1}^K A_k$.
\end{algorithm} 
}
After solving two subproblems, the overall algorithm to problem~(\ref{pb_o}) is summarized in Algorithm~\ref{alg:overall}. Notably, Algorithm~\ref{alg:overall} is an ADMM-based distributed algorithm, where the two subproblems are solved at the IoT device and edge server, separately. In each iteration, we first employ Algorithm~\ref{alg:single} to solve problem~(\ref{pb_o3}) with the computational complexity being 
$\mathcal{O}(F^{\mathrm{s}, \max}_k)$.
Thus, sequentially solving problem~(\ref{pb_o3}) results in the computational complexity of $\mathcal{O}(\sum_kF^{\mathrm{s}, \max}_k)$. 
Moreover, Algorithm~\ref{alg:single} can be executed for different IoT devices in parallel, and the computational complexity can be reduced.
After that, we obtain the optimal resource allocation strategy with Theorem~\ref{thm:resource}, and the computational complexity is $\mathcal{O}(K\mathrm{log} (\mu_{\max}))$. Note that only the required computation resource $\hat f_k^\mathrm{e}-\dfrac{\beta_k}{\rho}$ and the actual allocated resource $f_k^{\mathrm{e}}$ are exchanged between the IoT devices and the edge server.
Let $I_{\max}$ denote the maximum iteration number of Algorithm~\ref{alg:overall}, and the overall computational complexity of Algorithm~\ref{alg:overall} is $\mathcal{O}\left(I_{\max}\left(\sum_kF^{\mathrm{s}, \max}_k + K\mathrm{log} (\mu_{\max})\right)\right)$.
Additionally, we consider a limit algorithm: problem~(\ref{pb_o4}) can be solved using the interior point method with the computational complexity being $\mathcal{O}(K^{3.5}\mathrm{log}(1/\xi))$, where $\xi$ represents the precision of the solution. Therefore, the overall computational complexity of the limit algorithm
is $\mathcal{O}\left(I_{\max}(\sum_kF^{\mathrm{s}, \max}_k +K^{3.5}\mathrm{log}(1/\xi))\right)$. 
Compared to the proposed Algorithm~\ref{alg:overall}, one can clearly observe that the proposed algorithm exhibits a lower computational complexity.

To analyze the convergence of the ADMM-based distributed algorithm, we need to consider the convergence properties of its subproblems and the overall ADMM framework. Since we address problem~(\ref{pb_o3}) by performing the exhaustive search method, calculating the accuracy under each sampling rate, the convergence of Algorithm~\ref{alg:single} can be guaranteed. Algorithm~\ref{alg:KKT} solves a convex optimization problem using bisection search, which ensures convergence. Furthermore, according to the theoretical results in~\cite{wang2019global}, the global convergence of ADMM in nonconvex optimization can be guaranteed if the objective function satisfies certain properties, i.e., coercivity~\cite{1997nonlinear} and Lipschitz continuity. Given that the feasible domain of the objective function is bounded and $\alpha_k(F^{\mathrm{s}}_k,\tau_k)$ is locally Lipschitz continuous, these conditions are satisfied. Therefore, the proposed Algorithm~\ref{alg:overall} is guaranteed to converge.

To illustrate the practical deployment of the proposed algorithm, we describe its execution process within the multi-device IoT system.
In practical system operation, when the IoT device detects the onset of an action, it sends a sensing task request to the edge server. To enhance the efficiency of the multi-device
system, the edge server
employs a batch processing mechanism, triggering algorithm execution based on accumulated requests or waiting time~\cite{fowler2022survey}. Upon activation, Algorithm~\ref{alg:overall} performs the optimal resource allocation by considering updated factors such as current channel conditions, device power constraints, and sensing power levels. Once the algorithm converges, the IoT devices collect and upload sensing data, and the edge server executes the corresponding sensing tasks with the allocated resources. After all tasks are completed, the system remains idle until the next batch processing condition is met.

\section{Test Results}\label{sec:test}
In this section, we conduct experiments to demonstrate the effectiveness of the proposed framework and algorithm. 

\subsection{Test Setup}\label{subsec:test}

\begin{figure}[t]
	\setlength{\abovecaptionskip}{8pt} 	
	\centering
	\includegraphics[width=0.85\linewidth]{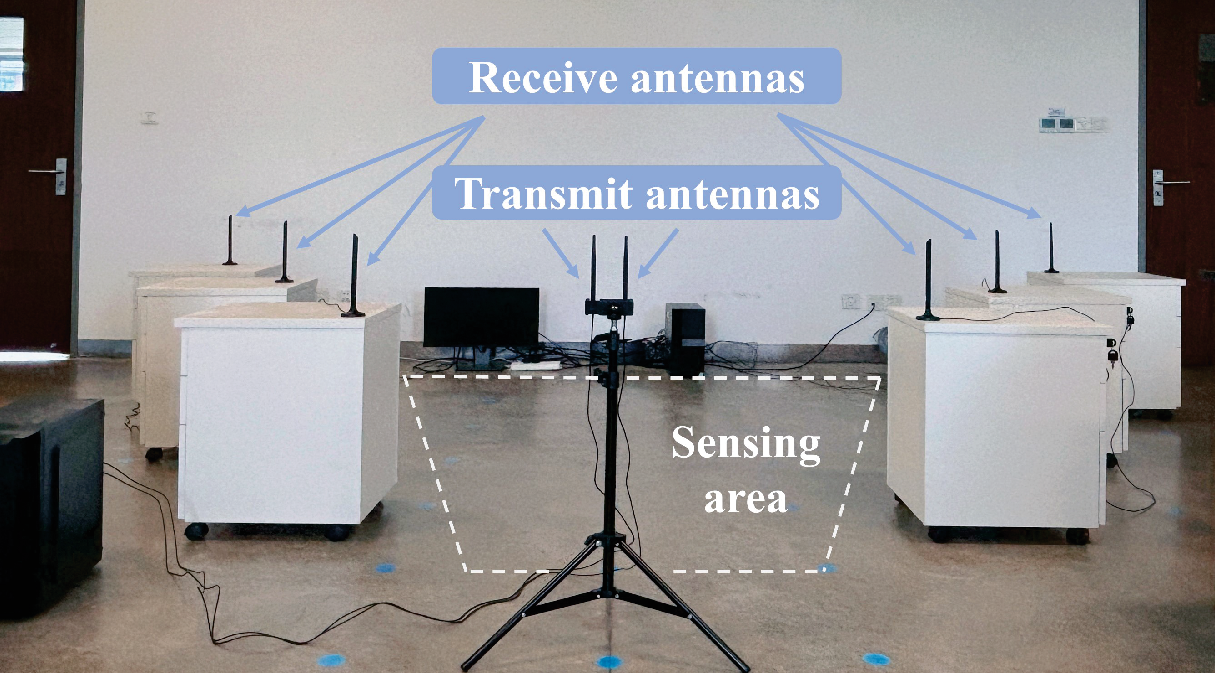}
	\vspace{-2ex}
	\caption{Real-world experiment scenario.}
	\label{fig:sce-real}
 \vspace{-2ex}
\end{figure}

Since our framework focuses on detecting the onset of movement, which requires continuous sensing data capturing the transition from static to active states, we first collect a custom dataset tailored to this specific requirement.
Specifically,
we conduct experiments in a typical indoor environment, an office room with a size of 10~\!m$\times$5~\!m.
We use one personal computer (PC) equipped with a two-antenna AX210 network interface controller (NIC) as the transmitter, and the other PC equipped with three AX210 NICs as receivers. The antennas of the transmitter and receivers are placed at different parts of the office, as shown in Fig.~\ref{fig:sce-real}. Moreover, both the transmitter and receivers operate at 5~\!GHz frequency band with a bandwidth of 20~\!MHz under the 802.11ac standard, and PicoScenes~\cite{jiang2021eliminating} is used to measure the CSI between the transmitter and receivers under the CSI sampling rate being 2,000~\!Hz. To collect the sensing dataset, eight volunteers are invited to perform eight daily activities, i.e., waving, kicking, bending down, raising a hand, walking, sitting down, standing up, and standing still, where standing still represents the static state. Each activity is performed 50 times for each person, resulting in 3,200 CSI time series in the dataset. The size of each CSI sample in the CSI time series is $12\times57$ with 12 being the number of transmitter-receiver antenna pairs and 57 being the number of available subcarriers. To simulate IoT devices at different locations with different allocated resource blocks, we randomly select CSI of one antenna pair and 10 subcarriers to construct the CSI sample and the corresponding dataset for each IoT device. Moreover, we adopt a CNN model with four convolution layers of 3 × 3 kernel size~\cite{cnn},
and it is trained on a Linux server equipped with four NVIDIA GeForce GTX 3080 GPUs. 

\begin{figure}[t]
	\setlength{\abovecaptionskip}{8pt} 	
	\centering
    \subfigure[Mean.]{
		\centering
		\includegraphics[width=0.7\linewidth, trim=3 0 2 20,clip]{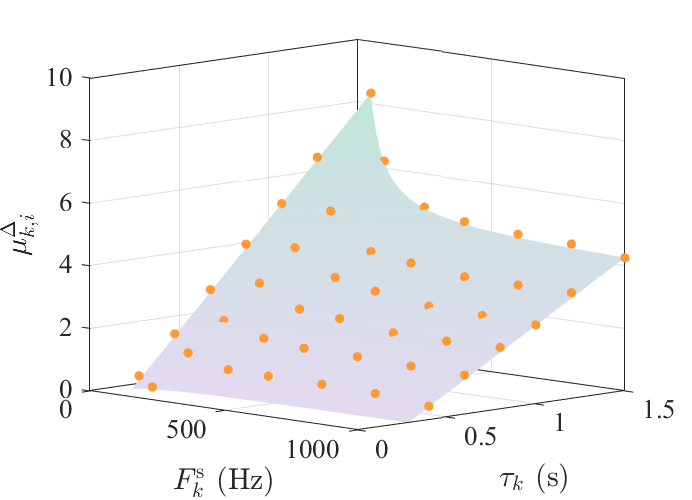}
	}
    
    \subfigure[Variance.]{
		\centering
		\includegraphics[width=0.7\linewidth,trim=3 0 2 20,clip]{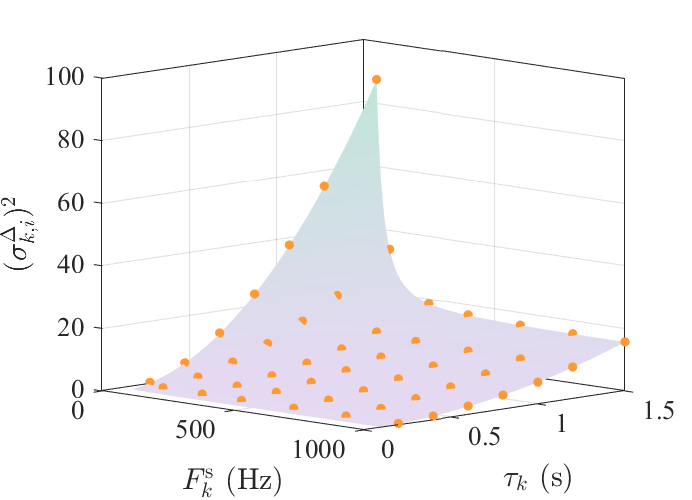}
	}
	\vspace{-2ex}
	\caption{Validation of the sensing model for the power difference of high-frequency components: example of ``raising a hand''.}
	\label{fig:mu_sigma}
\vspace{-2ex}
\end{figure}

After collecting the dataset and training the CNN for each device, we now simulate the sensing framework. We consider a BS with a radius of 500~\!m and 12 IoT devices are randomly located in the coverage.
The sensing delay requirement is set as 0.55~\!s for all devices.
For the local computation, the overall power upper limit and local computation resource for each IoT device follow the uniform distribution with $P_k^{\mathrm{over}} \in [26,29]$~\!dBm and $f^\mathrm{L}_k \in [35,50]$~\!MHz, respectively. Parameters $\lambda_i$, $r_i$, $\sigma^2_{\mathrm{c}}$, and $\sigma^2_{\mathrm{d},i}$ mentioned in Section~\ref{sec:sensing} are obtained via fitting on the collected dataset.
The requirement $p^{\min}$ on the miss rate and false positive rate is 2~\!\%.
For the wireless channel $h_{k,n}$ between each device and the BS, the large-scale fading $h_{k,n}$ is generated following the path loss model: $128.1+37.6\log_{10}(d)$, where $d$ denotes the distance between the device and the BS (in kilometer), and the small-scale fading follows Rayleigh distribution
with uniform variance. The transmit power for each device is 24~\!dBm. The bandwidth for communication is set as 4~\!MHz and the noise power $\sigma^2$ is set as -174~\!dBm/Hz. A Linux PC equipped with an Intel i9-13900K CPU is regarded as the edge server, and the total computation resource is 42~\!GHz. The accuracy function $\alpha_k(F^{\mathrm{s}}_k,\tau_k)$ is obtained using the trained CNN. Moreover, the probability for the static state is set as 0.4, and the probability for each of the remaining seven action types is therefore 0.6/7.

\begin{figure}[t]
	\setlength{\abovecaptionskip}{8pt} 	
	\centering
        \hspace{-3ex}
	\includegraphics[width=0.65\linewidth]{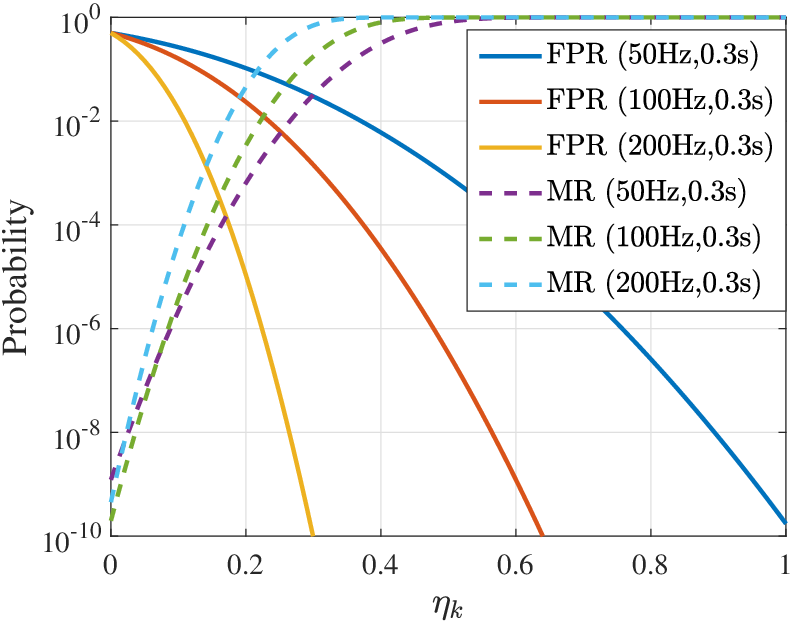}
	\vspace{-2ex}
	\caption{Miss rate (MR) and false positive rate (FPR) vs. threshold.}
	\label{fig:rate_eta}
    \vspace{-2ex}
\end{figure}

\begin{figure}[t]
	\setlength{\abovecaptionskip}{8pt} 	
	\centering
	\includegraphics[width=0.65\linewidth]{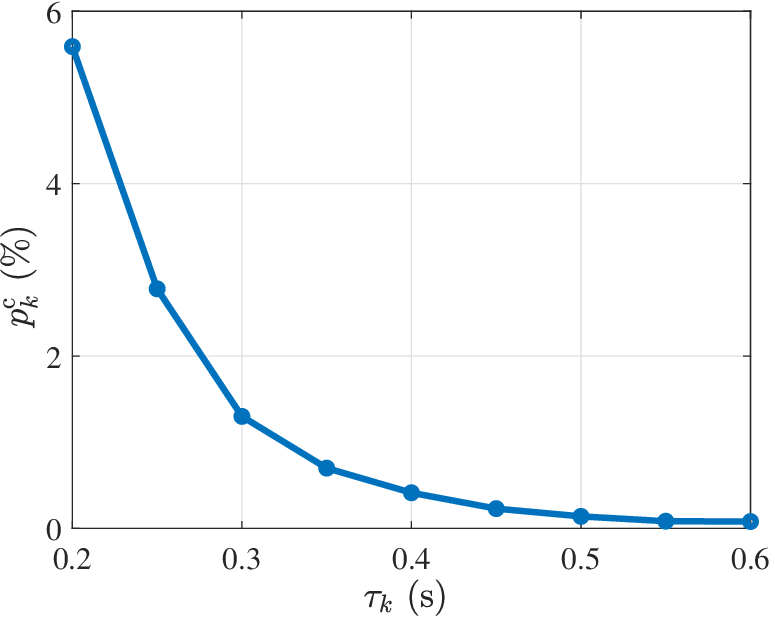}
	\vspace{-2ex}
	\caption{The intersection with different $\tau_k$ under $F_k^{\mathrm{s}}=100$~\!Hz.}
	\label{fig:p_c}
    \vspace{-2ex}
\end{figure}

\subsection{Verification for the Proposed Sensing Model}
We first pay attention to the proposed action detection module. To analyze its performance, we establish
the mathematical model for the power difference of high-frequency components, i.e., $\Delta P_k[l]$. To validate the 
correctness of the proposed model, we conduct tests on collected sensing data. Tab.~\ref{table: nmse} presents the normalized mean squared error (NMSE) between the real and fitted values of the mean and variance of $\Delta P_k[l]$ for eight types under different sample rates and time steps.
Specifically, the real values are directly calculated using the collected data, while the fitted values are obtained after fitting with~(\ref{eq:mu_static}),~(\ref{eq:mu_active_mean}), and~(\ref{eq:sigma_active_mean}).
It can be seen that fitting errors for all types are small ($\mathrm{NMSE}<0.1$), demonstrating the effectiveness of the established sensing model. Besides, we plot the mean and variance for one type (i.e., raising a hand) and the corresponding fitted functions under different sampling rates and time steps in Fig.~\ref{fig:mu_sigma}, with the remaining types showing similar results. It is evident that both the mean and variance decrease with the sampling rate while increase with the time step, and the fitted functions align well with the real results.

\begin{table}[b]
\vspace{-4ex}
	\centering
	\caption{NMSE between the real and fitted values.}
 \vspace{-1ex}
	\label{table: nmse}
 \resizebox{\linewidth}{!}{
	\begin{tabular}{|c|c|c|c|c|}
		\hline
		\!\!Type\!\!& Standing still& Kicking&Raising a hand & Waving  \\ 
       \hline
	Mean&$1.3 \times 10^{\!-2}$& $4.5 \times 10^{\!-3}$ &$2.3 \times 10^{\!-3}$&$3.1 \times 10^{\!-3}$\\
  \hline
	Variance&$1.2 \times 10^{\!-2}$&$2.5 \times 10^{\!-3}$&$6 \times 10^{\!-4}$&$7 \times 10^{\!-4}$\\
  
  \hline
  \!\!Type\!\!&Bending down& Walking  &Sitting down  & Standing up\\
  \hline
	Mean&$4.8 \times 10^{\!-3}$&$3.5 \times 10^{\!-3}$&$3.6 \times 10^{\!-3}$&$2.6 \times 10^{\!-3}$\\
 \hline
 Variance&$1.8 \times 10^{\!-3}$&$2.5 \times 10^{\!-3}$&$8 \times 10^{\!-4}$&$1.1 \times 10^{\!-3}$\\
 \hline 
	\end{tabular}}
\end{table}

To further explore the relationship among the miss rate, false positive rate, sampling rate, and time step, we plot the miss rate and false positive rate curves in Fig.~\ref{fig:rate_eta}. Here, we still take the action of ``raising a hand'' for example. From the figure, we can observe that the false positive rate decreases while the miss rate increases with the threshold, which verifies Proposition~\ref{prop:rate}. 
Besides, Fig.~\ref{fig:p_c} shows the intersection of the miss rate and false positive rate under a fixed sampling rate ($F_k^{\mathrm{s}}=100$~\!Hz). One can clearly observe that the intersection decreases with the time step, which verifies Theorem~\ref{thm:p}. 
Furthermore, with the results shown in the two figures, we can observe that, to reduce the miss rate and false positive rate at the intersection point, we need to increase at least one of the time step or sampling rate. The reason behind this is that a higher sampling rate or a wider time step allows more information to be collected, thus enhancing the performance of the action detection module.

\begin{figure}[t]
	\setlength{\abovecaptionskip}{8pt} 	
	\centering
	\includegraphics[width=0.65\linewidth]{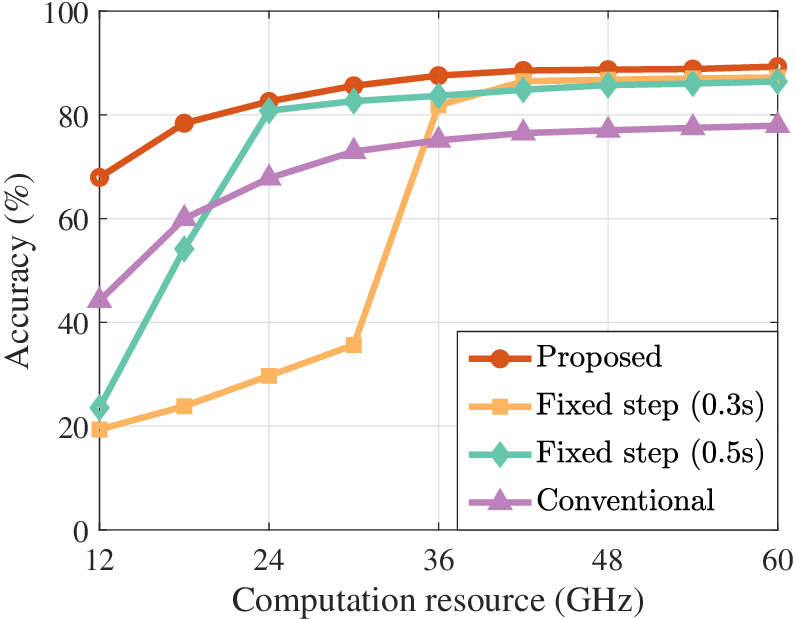}
	\vspace{-2ex}
	\caption{Accuracy vs. total edge computation resource.}
	\label{fig:acc_re}
    \vspace{-2ex}
\end{figure}

\subsection{Performance Comparison}

To demonstrate the superiority
of the proposed sensing framework and the corresponding algorithm, we consider two following baselines.

 $\bullet$ Conventional scheme~\cite{he2023integrated}: After collecting sensing data, each IoT device does not detect the onset of the sensing target's action. Instead, all sensing signals are uploaded to the edge server. After that, the edge server first detects the action by comparing the power of the high-frequency components with a threshold and then employs the CNN module to realize the action recognition.

$\bullet$ Fixed time step scheme: A similar sensing framework is adopted in this scheme, and the difference is that the time step is fixed. We consider two typical time steps: $0.3$~\!s and $0.5$~\!s. 
Other parameters are optimized using the same manner as the proposed algorithm.

First of all, we show the impact of the total edge computation resource on the sensing accuracy for three different schemes in Fig.~\ref{fig:acc_re}. 
It is readily seen that the accuracy of all schemes increases with the total edge computation resource. This is because the computation
delay of the CNN module in~(\ref{eq:com_cnn}) is positively correlated with the sampling rate and negatively correlated with the allocated edge computation resource. As the total computation resource increases, the allocated resource for each device also increases, which reduces the computation delay and enables a higher sampling rate that still satisfies the delay constraint. Besides, our proposed sensing framework exhibits a distinct performance improvement compared to the conventional scheme, particularly when the edge computation resource is insufficient. In the conventional scheme, continuous data transmission significantly increases the transmission delay, which in turn lowers the sampling rate and impairs the accuracy.
Moreover, compared to the fixed time step scheme, our proposal
shows better performance
because it adaptively adjusts the time step according to the dynamic channel.
The reduction in the computation resource leads to a drop in the sampling rate. To meet the constraints on the false positive rate and miss rate, the required time step should increase accordingly, indicated by Figs.~\ref{fig:rate_eta} and~\ref{fig:p_c}. In the fixed time step scheme, some devices fail to meet these constraints, thus leading to a sharp drop in accuracy.

\begin{figure}[t]
	\setlength{\abovecaptionskip}{8pt} 	
	\centering
	\includegraphics[width=0.65\linewidth]{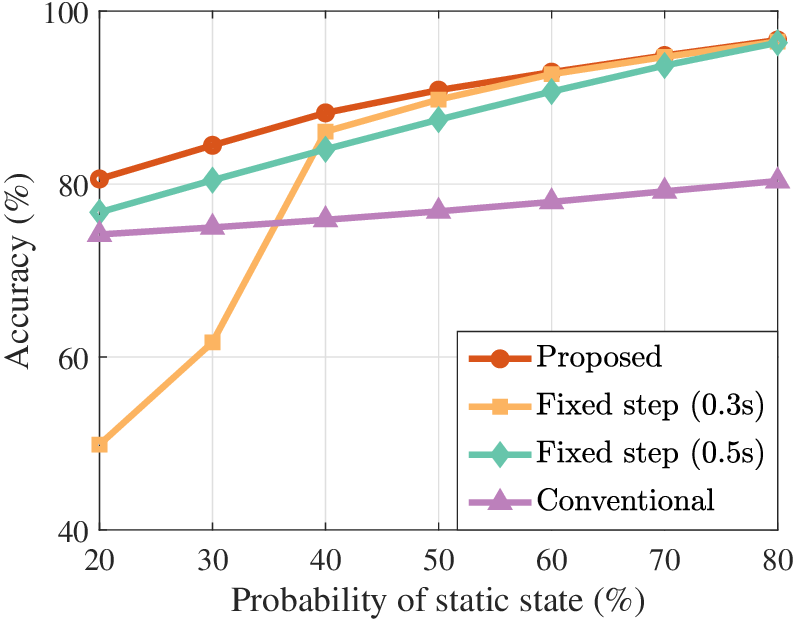}
	\vspace{-2ex}
	\caption{Accuracy vs. probability of the static state.}
	\label{fig:acc_p}
    \vspace{-2ex}
\end{figure}

Fig.~\ref{fig:acc_p} demonstrates the impact of the probability of the static state (i.e., the first phase in Fig.~\ref{fig:en_diff scenario}) on the sensing accuracy for three schemes. We can observe that the accuracy of the proposed scheme increases with the probability of the static state at a faster rate than the conventional scheme. 
The reason behind it can be explained as follows. When the probability of the static state is low, more time windows are filled with signals of interest and need to be transmitted to the edge server. Thus, the impact of the action detection module is reduced, and the required communication and computation resource of the proposed scheme are close to the conventional scheme. 
As the probability of the static state increases, fewer
time windows need to be transmitted, and the action detection module shows its effectiveness in saving communication and computation resources. In this case, a higher sampling rate can be achieved under the
delay constraint, 
thus increasing the sensing accuracy.
The accuracy of the fixed time step scheme also increases with the probability of the static state for the same reason mentioned before. However, when the static probability falls below 40\%, the accuracy of the fixed time step scheme ($\tau_k$=0.3~\!s) declines rapidly, further highlighting the necessity
of optimizing the time step.

\begin{figure}[t]
	\setlength{\abovecaptionskip}{8pt} 	
	\centering
	\includegraphics[width=0.65\linewidth]{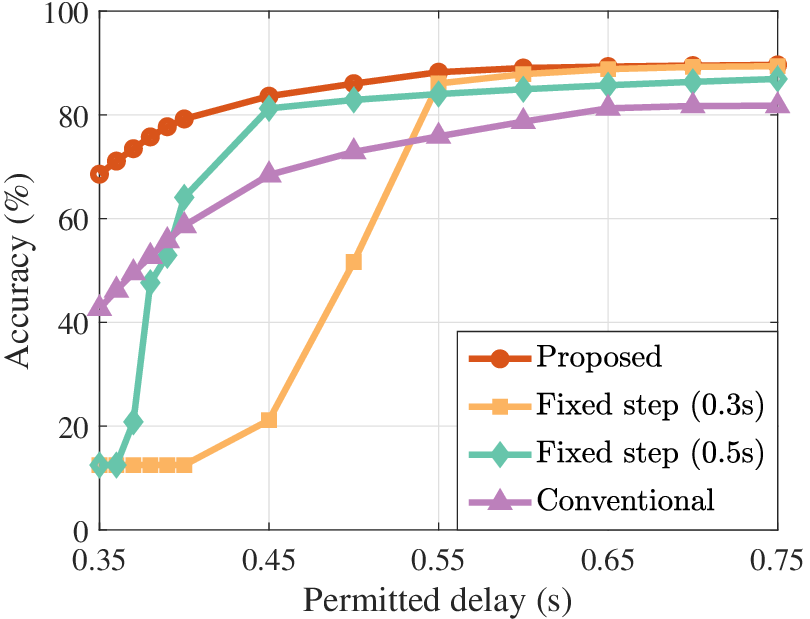}
	\vspace{-2ex}
	\caption{Accuracy vs. permitted delay.}
	\label{fig:acc_T}
    \vspace{-2ex}
\end{figure}

Fig.~\ref{fig:acc_T} depicts the impact of the permitted delay of sensing tasks on the sensing accuracy for three schemes.
Again, our proposed scheme demonstrates the best performance under varying permitted delays. Besides, it can be seen that the accuracy of all three schemes declines rapidly when the permitted delay is tight.
Such a phenomenon is caused by the sharp decline in the accuracy of the CNN module when the sampling rate is below 50~\!Hz, indicated by Fig.~\ref{fig:acc_F_initial}. 
Moreover, the sampling rate decreases as the permitted delay decreases, leading to the increase of the time step to meet the constraints on the false positive rate and miss rate. The combined effect of the decreased sampling rate and increased time step leads to a great decrease in the accuracy.
When the permitted delay is loose, the impact of the delay constraints becomes tiny. In this case, increasing the permitted delay leads to a higher sampling rate but contributes only slightly to the accuracy, since the accuracy of the CNN module grows slowly with the sampling rate exceeding 50~\!Hz.
Furthermore, compared with the two baselines, the sensing performance of our scheme is more stable under different permitted delays, demonstrating the robustness of our proposed scheme.

\begin{figure}[t]
	\setlength{\abovecaptionskip}{8pt} 	
	\centering
	\includegraphics[width=0.65\linewidth]{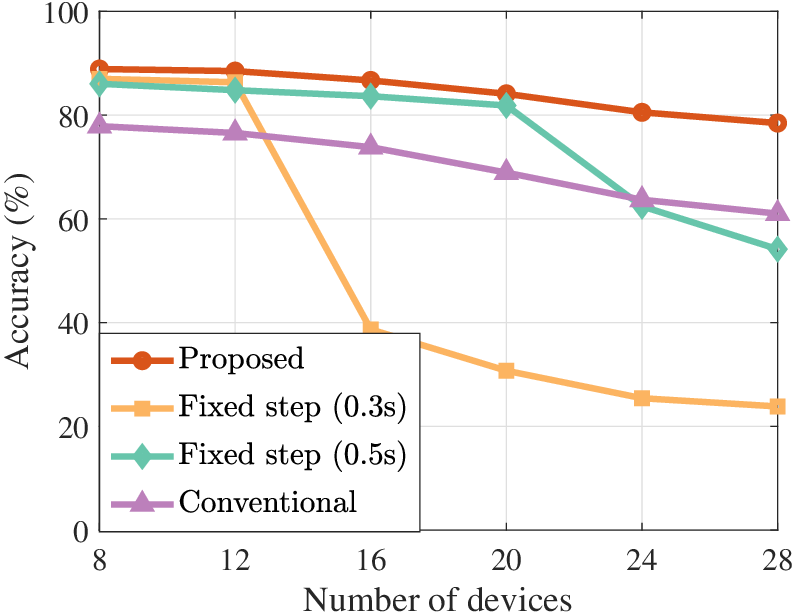}
	\vspace{-2ex}
	\caption{Accuracy vs. number of devices.}
	\label{fig:acc_N}
    \vspace{-2ex}
\end{figure}

Fig.~\ref{fig:acc_N} shows the impact of the number of devices on the sensing accuracy for three different schemes. Here, we simulate an increased number of devices by randomly selecting data from the existing 12 devices to construct additional device datasets.
From the figure, the performance of all schemes degrades with the number of devices. It is because devices compete for the limited edge computation resource, and thus, each device is allocated less resource as the number of devices increases. With less resource, the computation delay increases, resulting in a lower sampling rate to meet the delay constraint.
When the number of devices is less than 12, the edge computation resource is sufficient, and all schemes can achieve high sampling rates, leading to marginal variation in accuracy under different numbers of devices.
Besides, the accuracy of the proposed scheme declines the slowest among all schemes, again demonstrating the superiority of our proposed scheme.

\section{Conclusion} \label{sec:Conclusion}

In this paper, we studied
a multi-device ISCC system and proposed a resource-efficient sensing framework, including
a novel action detection module deployed at each IoT device and an edge recognition module deployed at the BS. 
With the ability to identify the onset of an action, the action detection module effectively reduces unnecessary data transmission and computation. 
Mathematical models were established to quantitatively analyze the sensing performance of the proposed sensing framework.
To further improve the sensing performance in the multi-device ISCC system, we formulated
a sensing accuracy maximization problem 
considering the power consumption, the delay requirements of sensing tasks, and the computation resource limitations at the edge server. Subsequently, we proposed an ADMM-based distributed algorithm, decomposing the original problem into two subproblems and solving them separately at the 
device and edge server.
Finally, we conducted a real-world test to validate the proposed sensing model and demonstrate
the superiority of the proposed sensing framework and algorithm.

While the proposed sensing framework has shown promising results, some problems still require further investigation. First, in this paper, we adopted a time-division manner for joint communication and sensing, ensuring compatibility with existing IoT systems. However, resource utilization could be further improved by leveraging shared time resources. A feasible approach is to use existing dual-functional waveforms for sensing and communication simultaneously. Second, in scenarios with a large number of devices, potential challenges such as spectrum resource scarcity and network congestion, may arise. To address these, an intelligent resource block scheduling strategy should be designed, dynamically adapting to real-time traffic demand and channel conditions. Third, the proposed sensing framework can be extended to multi-modal sensing scenarios, where aggregation techniques could be incorporated into the action detection module to further enhance the overall performance. Finally, our current framework assumed the presence of a direct path between IoT devices and the targets. However, in practical environments, the direct path may be blocked by obstacles. A promising solution to this issue is to leverage reconfigurable intelligent surfaces to configure exclusive paths for sensing~\cite{yu2023active,wan2024reconfigurable}.

\appendices
\section{Proof of Theorem~\ref{thm:p}}\label{proof:p}
First of all, we can rewrite $p_{k,i}^{\mathrm{o}}$ and $p_{k}^{\mathrm{l}}$ as
\begin{align} p_{k,i}^{\mathrm{o}}\!&=\!Q\!\left(\frac{\left(\frac{\lambda_i-\lambda_1}{2T^{\mathrm{s}}}+\frac{r_i-r_1}{ 2(T^{\mathrm{s}})^2F^{\mathrm{s}}_k}\right)-\eta_k/\tau_k}{\sqrt{\!\left(
    \frac{4\sigma^2_{\mathrm{c}}(\lambda_i\!+\!4\lambda_1)}{3(T^{\mathrm{s}})^3F^{\mathrm{s}}_k}\! 
   \! +\! \frac{2\sigma^2_{\mathrm{c}}(r_i\!+\!4r_1)}{3(T^{\mathrm{s}})^4 (F^{\mathrm{s}}_k )^2}
  \right)\!\!+\!\frac{\sigma^2_{\mathrm{d},i}}{\tau_k^2}}}\!\!\right)\!,\label{eqa:p1}\\
p_{k}^{\mathrm{l}}\!&=\!\!Q\!\left(\frac{\eta_k/\tau_k}{\sqrt{\left( \frac{8\sigma^2_{\mathrm{c}}}{(T^{\mathrm{s}})^3F^{\mathrm{s}}_k}\lambda_1  +  \frac{4\sigma^2_{\mathrm{c}} }{(T^{\mathrm{s}})^4 (F^{\mathrm{s}}_k )^2}r_1\right)\!+\!\frac{\sigma^2_{\mathrm{d},1}}{\tau_k^2}}}\right)\!,\label{eqa:p2}
\end{align}
respectively. Due to the complexity of the above expressions, it is hard to directly prove that $p^{\mathrm{c}}_k$ decreases with $\tau_k$. Instead, we opt to prove that $p_{k,i}^{\mathrm{o}}$ decreases with $\tau_k$ under the same $p_{k}^{\mathrm{l}}$, which can also prove the theorem. To this end, we first give the condition for maintaining the same $p_{k}^{\mathrm{l}}$ under varying $\tau_k$, as 
\begin{equation}
    \eta_k/\tau_k=\omega \sqrt{\left( \dfrac{8\sigma^2_{\mathrm{c}}}{(T^{\mathrm{s}})^3F^{\mathrm{s}}_k}\lambda_1  +  \dfrac{4\sigma^2_{\mathrm{c}} }{(T^{\mathrm{s}})^4 (F^{\mathrm{s}}_k )^2}r_1\right)\!+\!\dfrac{\sigma^2_{\mathrm{d},1}}{\tau_k^2}},
\end{equation}
where $\omega$ is a constant. From this, we can observe that $\eta_k/\tau_k$ decreases with $\tau_k$. By substituting this into the expression (\ref{eqa:p1}) for $p_{k,i}^{\mathrm{o}}$, we can find that the numerator within the Q-function increases, while the denominator decreases with $\tau_k$. Since the Q-function is a monotonically decreasing function, $p_{k,i}^{\mathrm{o}}$ decreases with $\tau_k$. 

Based on the above analysis, we can conclude that under the same $p_{k}^{\mathrm{l}}$, $p_{k,i}^{\mathrm{o}}$ decreases with $\tau_k$. As a result, the intersection point $p^\mathrm{c}_{k}$ also decreases with $\tau_k$, which ends the proof.

\vspace{-1ex}
\section{Proof of Theorem~\ref{thm:resource}} \label{proof:resource}
According to the KKT conditions, the necessary and sufficient conditions for the optimal solution can be formulated as follows
\begin{align}
	\frac{\partial \mathcal{L}}{\partial f_k^{\mathrm{e},\star}} & = \frac{\rho}{K}(f_k^{\mathrm{e},\star}\!-\!\hat f_k^\mathrm{e}\!+\!\frac{\beta_k}{\rho})\!+\!\mu^{\star}\! \!\left\{
	\begin{array}{ll}
		\!=0, &\!\!\!f_k^{\mathrm{e},\star} > 0,\\
		\!\ge 0,&\!\!\!f_k^{\mathrm{e},\star} = 0,
	\end{array}	
	\right. \label{kkt-1}
\end{align}
\vspace{-2ex}
\begin{align}
    \mu^{\star} (\sum_{k=1}^{K}f_k^{\mathrm{e},\star}- f^{\mathrm{e}})=0,~~\mu^{\star} \ge 0.  \label{kkt-2}
\end{align}
Note that there are two cases for $\mu^{\star}$.
\begin{itemize}
    \item When $\mu^{\star}=0$, $\sum_{k=1}^{K}f_k^{\mathrm{e},\star}\le  f^{\mathrm{e}}$. According to~(\ref{kkt-1}), 
    $f_k^{\mathrm{e},\star}=\hat f_k^\mathrm{e}-\dfrac{\beta_k}{\rho}$ if $f_k^{\mathrm{e},\star}>0$; otherwise, $f_k^{\mathrm{e},\star}$ should be zero. Thus, in this case, $f_k^{\mathrm{e},\star}$ can be expressed as $f_k^{\mathrm{e},\star}=\max \{\hat f_k^\mathrm{e}-\dfrac{\beta_k}{\rho},0\}$.
    \item When $\mu^{\star}>0$, $\sum_{k=1}^{K}f_k^{\mathrm{e},\star}= f^{\mathrm{e}}$. Similarly, $f_k^{\mathrm{e},\star}$ can be expressed as $f_k^{\mathrm{e},\star}=\max \{\hat f_k^\mathrm{e}-\dfrac{\beta_k}{\rho}-\dfrac{K}{\rho}\mu^{\star},0\}$.
\end{itemize}
Thus, we can derive the optimal solution as shown in Theorem~\ref{thm:resource}, which ends the proof.

\bibliographystyle{IEEEtran}
\bibliography{ref}

\end{document}